\documentclass[11pt,a4paper]{article}
\pdfoutput=1
\usepackage{jheppub}

\usepackage{tikz}
\usetikzlibrary{arrows.meta, positioning}

\usepackage{amsmath}
\usepackage{amsfonts}
\usepackage{verbatim}
\usepackage{graphicx}
\usepackage{mathrsfs}
\usepackage{appendix}
\usepackage{caption}
\usepackage{float}
\usepackage{subfig}
\usepackage{mathtools}
\usepackage{mathtools}
\usepackage{xcolor}
\usepackage{comment}
\usepackage{gensymb}

\usepackage{colortbl}
\usepackage{xcolor}

\definecolor{carr1}{RGB}{173, 216, 230} 
\definecolor{carr2}{RGB}{144, 238, 144} 
\definecolor{carr12}{RGB}{200, 180, 220} 

\newcommand{\gb}{\hat{\gamma}}

\newcommand{\gc}{\tilde{\gamma}}

\newcommand{\p}{\partial}

\newcommand{\g}{\gamma}

\newcommand{\s}{\sigma}

\newcommand{\rw}{\rightarrow}

\title{Carroll fermions, expansions and the lightcone}
\author{Arjun Bagchi}\author{and Saikat Mondal.} 
\affiliation{Indian Institute of Technology Kanpur, Kanpur 208016, India.}
\emailAdd{abagchi@iitk.ac.in}
\emailAdd{saikatmd@iitk.ac.in}

\abstract{We investigate fermions on Carrollian manifolds. We complement previous intrinsic analysis by deriving Carrollian fermion actions from a relativistic Dirac theory via a systematic expansion in the speed of light ($c$). We then study relativistic fermions in light-cone coordinates and their connection to Carrollian fermions in one lower dimension. This follows from the recent observation that the Poincaré algebra, written in lightcone coordinates contains (two) co-dimension one Carroll sub-algebras. Our results establish a clear bridge between intrinsic Carrollian constructions, small $c$-expansion and light-cone dynamics. In the process, we understand why  Carrollian fermions in $D$-dimensions have features that relate them to relativistic fermions in both $D$ and $(D+1)$ dimensions.}

\begin{document}

\maketitle

\preprint{}


\section{Introduction}\label{sec1}
Fermions are of central importance to all of physics since they make up all of visible matter in the present day universe. As the fundamental degrees of freedom in our universe are relativistic, relativistic fermions play a central role in conventional quantum field theory with wide-ranging applications from high-energy physics to condensed matter systems. The dynamics of relativistic fermions is dictated by Lorentz symmetry which constrains both kinematics and allowed interactions. 

\medskip

However, in many physically relevant scenarios -- most familiar of which are in real life condensed matter physics and non-relativistic hydrodynamics -- we are faced with the situation where the assumption of Lorentz symmetry becomes inadequate. In these situations, Galilean symmetry replaces Lorentzian invariance as the symmetry underlying these physical systems. One is thus led to symmetries beyond Lorentzian symmetries and structures beyond the usual pseudo-Riemannian manifolds that come hand-in-hand with Lorentzian symmetries. We will be interested in these {\emph{non-Lorentzian}} set-ups and particularly in a systematic investigation of fermionic field theories in non-Lorentzian geometries. 

\medskip

Apart from the above mentioned Galilean symmetries, where one sends the speed of light $c\to \infty$, in recent years, Carrollian symmetry has emerged as a particularly promising organisation principle. Here the speed of light is sent to zero $(c\to 0)$ \cite{LevyLeblond, SenGupta:1966qer} instead of infinity and consequently the light-cones close up and ultra-locality emerges. 

\medskip

Carrollian symmetry has attracted renewed interest due to its appearance in diverse physical settings, most notable among which is its isomorphism with the Bondi-van der Burg-Metzner-Sachs (BMS) symmetries \cite{Bondi:1962px, Sachs:1962wk, Ashtekar:1996cd, Barnich:2006av} that are asymptotic symmetries at null infinity of asymptotically flat spacetimes \cite{Duval:2014lpa, Duval:2014uva}.  This has led to the whole programme of Carrollian holography that is one of the main avenues of constructing holography for asymptotically flat spacetimes \cite{Bagchi:2010eg, Bagchi:2012xr, Barnich:2012xq, Bagchi:2014iea, Bagchi:2015wna, Bagchi:2016bcd, Donnay:2022aba, Bagchi:2022emh, Bagchi:2023fbj, Saha:2023hsl, Bagchi:2023cen, Alday:2024yyj}. Carrollian symmetries also appear on the horizons of black holes \cite{Penna:2018gfx,Donnay:2019jiz} as well as in near horizon regions \cite{Bagchi:2026qpi}; as residual gauge symmetries on the worldsheet of tensionless strings \cite{Bagchi:2015nca,Bagchi:2016yyf, Bagchi:2026wcu}; in condensed matter systems \cite{Bidussi:2021nmp, Bagchi:2022eui, Biswas:2025dte} and in ultra-relativistic hydrodynamics \cite{Bagchi:2023ysc, Bagchi:2023rwd}. For a comprehensive review of this rapidly developing field, we refer the reader to \cite{Bagchi:2025vri} and references therein. 

\medskip

One of the reasons behind this all-pervading nature of Carrollian symmetries is that Carrollian structures naturally appear on all null surfaces and this makes Carrollian structures a unifying language for physics on null hypersurfaces. We refer the reader to \cite{Bergshoeff:2022eog, Ciambelli:2025unn} for a thorough geometric guide to Carroll symmetries. It is particularly interesting to note that when written in lightcone coordinates $(x^\pm, x^i)$, the Poincaré group naturally contains two lower dimensional Carroll groups in it \cite{Bagchi:2024epw, Majumdar:2024rxg}, corresponding to the two null directions $x^+$ and $x^-$, in addition to the well-known lower dimensional Galilean groups \cite{Weinberg:1966jm, Susskind:1967rg, Kogut:1969xa} which have been instrumental in understanding the lightfront formulation of gauge theories, Discrete Light Cone Quantization and connections to M-theory \cite{Brodsky:1997de, Banks:1996vh, Susskind:1997cw}. 

\medskip

\subsection{Carrollian field theories, fermions and expansions}

Given the wide-ranging applications, the development of classical and quantum field theories with Carrollian symmetries is a crucial task. In the early years of their development, Carrollian symmetries were investigated at the level of equations of motion of the classical theories \cite{Bagchi:2016bcd, Bagchi:2019clu, Bagchi:2019xfx}. This later progressed to a more detailed study of scalar theories \cite{deBoer:2021jej, Henneaux:2021yzg, deBoer:2023fnj} and spin-1 gauge fields \cite{deBoer:2021jej, Henneaux:2021yzg, deBoer:2023fnj}. Carroll fermions were not investigated in detail until \cite{Bagchi:2022eui, Yu:2022bcp, Hao:2022xhq} \footnote{See however studies related to tensionless superstrings in \cite{Bagchi:2016yyf, Bagchi:2017cte} where similar structures were discussed. More aspects of 2D Carroll fermions were addressed in \cite{Banerjee:2022ocj}.}. There have since been some more work in this direction \cite{Bergshoeff:2023vfd, Ekiz:2025hdn}. However, some fundamental issues remained and we will address some of these in our paper. 

\medskip

As we will go on to explain in detail below, a defining feature of fermions in a Carroll setting is the fundamental change in the underlying Clifford algebra, which reflects the non-Lorentzian nature of the spacetime. The usual Lorentzian Clifford algebra is inseparably linked with the Minkowski metric
\begin{align}
    \{\gamma^\mu, \gamma^\nu \} = 2\eta^{\mu\nu}.
\end{align}
Here $\eta_{\mu\nu} = \text{diag}(-c^2, 1, \ldots 1)$. As one takes $c\to 0$, the metric degenerates, giving rise to Carrollian structures. 
\begin{align}
    \eta_{\mu\nu} \to h_{\mu\nu} = \text{diag}(0, 1, \ldots 1), \quad -c^2 \eta^{\mu\nu} \to \Theta^{\mu\nu} = \text{diag}(1, 0, \ldots 0) = \theta^\mu \theta^\nu. 
\end{align}
The Carroll Clifford algebra thus takes a fundamentally different form 
\begin{align}\label{CClf} \{\gc_{\mu},\gc_{\nu}\}=2h_{\mu\nu},~~\quad \{\gb^{\mu},\gb^{\nu}\}=2\Theta^{\mu\nu}. 
\end{align}
We see the emergence of two varieties of Carroll fermions based on these two different Clifford algebras. As we will explain in detail in our review of Carroll fermions below, the degenerate nature of the Carroll Clifford algebra leads to many stark differences with usual relativistic fermions.

A systematic route to constructing Carrollian theories is provided through an expansion in powers of speed of light, commonly referred to as $c$-expansion, of a parent relativistic theory. In this approach, relativistic actions, equations of motion, and symmetry transformations are expanded around the $c=0$ point, allowing Carrollian dynamics to emerge in a controlled order-by-order manner. The $c$-expansion has been successfully applied to scalar, gauge, and gravitational systems \cite{deBoer:2021jej,Hansen:2021fxi}, providing a systematic and controlled derivation of Carrollian field theories from relativistic dynamics. Carroll fermions have not been treated in this $c$-expansion previously, and we will address this problem below. Our analysis will throw up various interesting surprises. 

\medskip

The leading pieces in a $c$-expansion are manifestly Carroll invariant and these theories have been dubbed ``Electric'' theories, borrowing nomenclature from the non-relativistic world and specifically Galilean Electrodynamics \cite{LeBellac:1973unm, Bagchi:2014ysa, Festuccia:2016caf}. The sub-leading pieces typically break Carroll boost invariance, but symmetries can be restored by adding suitable Lagrange multipliers. The next-to-leading order (NLO) pieces make up the so called ``Magnetic'' Carroll theory. Given we have two types of Carroll fermions, an expectation was that these become electric and magnetic fermions in an expansion. We will see that this expectation is met, although with some interesting twists. 

\medskip

An important little detail to keep in mind is that in previous literature, which dealt exclusively with bosonic fields, the $c$ expansion was actually a $c^2$ expansion, i.e. only even powers were considered. It was expected that nothing substantially different would be achieved by considering odd powers of $c$ {\footnote{See \cite{VandenBleeken:2017rij} for a discussion of odd powers in the Galilean $1/c$ expansion.}}. We will see that fermions necessitate the considering of odd powers and it is down to the observation that the Clifford algebra has a metric on the right hand side which is expanded in $c^2$, so the gammas on left hand side have to expand in odd powers. Though we will not be expanding gammas in what follows, the logic with the fermionic fields is similar. 

\subsection{Oddity in odd dimensions and lessons from the lightcone}

We will consider many aspects of Carrollian fermions in our work. One of the peculiarities we will come across would be that the dimension of the representations of Carroll fermions in odd spacetime dimensions is very different from relativistic fermions in the same dimension. For example, we know that when spacetime dimension is $D=3$, the defining representation of relativistic fermions is 2d. We will show that in $D=3$, the degenerate nature of the Carroll Clifford algebra necessitates a 4d representation for Carroll fermions. This hints at something and a relationship between the $D=3$ Carroll group and the $D=4$ Poincaré group \footnote{A note about notations: we will use $D$ for spacetime dimensions and $d$ for the dimension of the representations of Clifford algebras.}.  

\medskip

As we mentioned earlier, the Poincaré group written in lightcone coordinates reveals very interesting sub-structures. It has been known since long \cite{ Weinberg:1966jm, Susskind:1967rg, Kogut:1969xa} that there was a co-dimension one Galilean subgroup which emerged out of this representation and this has been behind the formulation of lightcone field theories and their connections to non-relativistic systems. We now realise that there are also two co-dimension one Carroll subgroups associated with the lightcone Poincaré group, one corresponding to each null direction. Thus a different and physically intuitive realization of Carrollian structures is expected to arise in the light-cone formulation of relativistic field theory. This would give a lower dimensional Carrollian theory from a higher dimensional relativistic theory.  

\medskip

By introducing coordinates adapted to the lightcone, spacetime is naturally foliated by light-like slices, and the Lorentz (or Poincaré) algebra reorganizes into generators adapted to these null directions. Starting from the seminal work by Dirac \cite{Dirac:1949cp}, light-cone (or light-front) quantization has a long history as a Hamiltonian framework adapted to null hypersurfaces \cite{Weinberg:1966jm, Susskind:1967rg, PhysRevD.1.2901, PhysRev.180.1506}. In this formulation, the light-cone momentum plays the role of Hamiltonian, and the spinor field naturally splits into dynamical and constrained components -- a feature we will see is intimately connected to the Carroll reduction. For a comprehensive review of light-front dynamics, see \cite{Brodsky:1997de}. This reorganization leads to a nontrivial reduction of the relativistic symmetry structure, closely resembling the symmetry contraction underlying Carrollian spacetimes.

This null-adapted formulation has important consequences for fermionic dynamics. Fermionic representations differ qualitatively from their relativistic counterparts: some components of the Dirac spinor become non-dynamical, while the symmetry algebra governing the fermions resembles a non-Lorentzian (Carroll/Galilei) instead of the Lorentz algebra. Light-cone quantization may then be regarded as a quantization scheme adapted to this null foliation, making the reduced dynamical content of the theory manifest.

From this perspective, light-cone fermions can be viewed as furnishing a concrete realization of fermionic dynamics on Carrollian backgrounds. The reduction of the Lorentz algebra induced by the light-cone decomposition mirrors the symmetry reduction inherent to Carrollian spacetimes, as shown in \cite{Bagchi:2024epw, Majumdar:2024rxg}. Understanding how gamma matrix structures, spinor representations and fermionic equations of motion reorganize in this limit is therefore of both conceptual and practical importance. In particular, the reason for the fundamentally different behaviour of Carroll fermions in odd dimensions to their relativistic counterparts is manifest in terms of the lightcone. 

\paragraph{Outline of the paper:} The outline of the paper is as follows. In Section \ref{sec2}, we provide a brief review of Carrollian symmetry and the associated Carroll Clifford algebra, with particular emphasis on their representations in even and odd spacetime dimensions. Section \ref{sec3} is devoted to the $c-$expansion of the relativistic Dirac action, where the Carrollian fermionic theory emerges in the leading and sub-leading order. Section \ref{sec4} explores fermionic dynamics in a null-adapted setting, elucidating how null contractions and light-cone coordinates lead to a reorganization of the symmetry algebra and fermionic degrees of freedom.

\section{Carrollian fermions}\label{sec2}
\subsection{Carroll symmetries}
We begin by briefly recalling the geometric framework underlying Carrollian manifolds \cite{Henneaux:1979vn, Duval:2014uoa}. A Carroll manifold is a $D$-dimensional manifold equipped with the geometric doublet $\{h_{\mu\nu}, \theta^\rho\}$, where $h_{\mu\nu}$ is a symmetric covariant tensor  of rank $(D-1)$ and signature $(0, + \ldots +)$ interpreted as a spatial metric and $\theta^\mu$ is a nowhere-vanishing vector field spanning its kernel, 
\begin{align}
    h_{\mu\nu} \theta^\nu = 0. 
\end{align}
A flat Carroll structure can be realised by choosing 
\begin{eqnarray}\label{FlC}
h_{\mu\nu} = \begin{pmatrix}
			0 & 0 \\
			0 & \quad I_{D-1}
		\end{pmatrix}, 
		\qquad
\Theta^{\mu\nu} = 	\begin{pmatrix}
			1 & 0 \\
			0 & \quad 0_{D-1} 
		\end{pmatrix}   \, = \theta^\mu \theta^\nu.
\end{eqnarray}
The isometries preserving this flat Carrollian data form the Carroll group. The isometry equations
\begin{align}
    \pounds_{\xi}h_{\mu\nu}=0, ~~~\quad \pounds_{\xi}\theta^{\mu}=0 \,
\end{align}
yields the Carroll Killing vectors
\begin{align}\label{CarKill}
    \xi^i = \omega^i_{\,j}\, x^j + b^i ,\qquad \xi^0 = a + f(x^k)\,,
\end{align}
where $\omega^i_{\,j}$ generate spatial rotations, $b^i$ and $a$ correspond to spatial and time translations respectively and $f(x^k)$ is an arbitrary function of spatial coordinates, giving rise to the so-called infinite dimensional supertranslations. When the function $f(x^k)$ is restricted to be linear, one obtains a finite dimensional symmetry algebra generated by spatial rotations $J_{ij}$, spatial translations $P_i$, Carroll boosts $C_i$ and time translations $H$, whose non-vanishing commutators are given by
\begin{align}\label{carral}
        [J_{ij},J_{kl}]&=so(D-1), \quad [C_i,P_j]=-\delta_{ij}H, \quad [J_{ij},X_{k}]=-\delta_{ik}X_{j} + \delta_{jk}X_{i}.
\end{align}     
where $X_i = P_i, C_i$. This finite algebra can also be obtained via an Inönü-Wigner contraction of the Poincaré algebra by taking the speed of light $c \rightarrow 0$ limit \cite{LevyLeblond, SenGupta:1966qer}. At the level of spacetime coordinates, this corresponds to the scaling \cite{Bagchi:2012cy}
    \begin{equation}
		x^i\rw  x^i ,\quad t\rw \epsilon t, \quad \epsilon\rw 0\,.
    \end{equation}
In this limit, the Carroll generators admit the following realisation in position space:
\begin{equation}
H=\p_{t} ,\quad P_i=\p_{i} ,\quad C_i=x_i\p_{t} ,\quad J_{ij}=x_i\p_{j}-x_j\p_{i}.
\end{equation}
Distinguishing features of the Carroll algebra are that the boosts commute among themselves, while the Hamiltonian $H$ appears as a central element. The vanishing of the commutator between Carroll boosts and the Hamiltonian gives rise to the very typical {\em flat band} feature of Carrollian systems where the energy dispersion relation becomes trivial \cite{Bagchi:2022eui}. These structural differences relative to the Poincaré algebra lead to markedly different physical behaviour in the Carroll world. The above finite generators generate the Carroll transformation on space-time coordinates as
\begin{align}		
t{'}=t+a -\vec{v}.\vec{x}, \quad \vec{x}{'}=\boldsymbol{R}\vec{x}+\vec{b}\,.
\end{align}  
where $a, \vec{b},\vec{v}$, and $\boldsymbol{R} \in SO(D-1)$ parametrise time translations, spatial translations, Carroll boosts and rotations respectively. 
We notice that this is the time $\leftrightarrow$ space swapped version of Galilean transformations. The underlying Carrollian and Newton-Cartan manifolds also enjoy this duality in their fibre-bundle structure with the transition between Carroll and Galilei controlled by a base to fibre swap \cite{Duval:2014uoa}.

\subsection{Constructing Carroll fermions}
We now summarize the basic features of Carrollian fermions following \cite{Bagchi:2022eui} \footnote{For a detailed discussion particularly in 2D and related research, the reader is pointed to \cite{Banerjee:2022ocj, Hao:2022xhq, Yu:2022bcp}. An alternative limiting construction was initiated in \cite{Bergshoeff:2023vfd} and quantum aspects has been studied in \cite{Ekiz:2025hdn}}. As mentioned in the introduction, the degenerate structure of the Carrollian manifold necessitates a fundamental change in the Clifford algebra from its relativistic form and the Carrollian Clifford algebra is given by \eqref{CClf}. This structure naturally introduces two distinct sets of Carroll gamma matrices, together with their respective representations, leading to two inequivalent fermionic theories: the \textit{lower gamma} theory (action constructed from $\gc_{\mu}$) and the \textit{upper gamma} theory (action constructed from $\gb^{\mu}$). These capture the different possible ways fermions can behave in a Carrollian background. 

A defining requirement of the Carroll Clifford algebra is that the spacetime generators constructed from it close into the Carroll algebra, especially homogeneous part of it. For the lower gammas, we define the spin generators as 
\begin{align}\label{sig}
    {\Sigma}_{\mu\nu} = \frac{1}{4} [\gc_\mu, \gc_\nu].
\end{align}
One can easily verify that these generators indeed satisfy the claim i.e. 
\begin{align}
    [{\Sigma}_{\mu\nu}, {\Sigma}_{\rho\sigma}] = h_{\nu \rho}\Sigma_{\mu\sigma} - h_{\mu\rho}\Sigma_{\nu\sigma} + h_{\mu\sigma}\Sigma_{\nu\rho} - h_{\nu \sigma}\Sigma_{\mu\rho} \,,  
\end{align}
where $h_{00}=0$ specifically means $[{\Sigma}_{0i}, {\Sigma}_{0j}] = 0$, i.e. Carroll boosts commute. 
The generators constructed from the upper gamma matrices follow an analogous algebra with the metric $h_{\mu\nu}$ replaced by $\Theta^{\mu\nu}$.

We now turn to the construction of Carroll-invariant spinor bilinears. Starting from a spinor $\Psi$, our first task is to define an appropriate notion of adjoint. Owning to the degenerate nature of the Carroll Clifford algebra, the usual definition of the adjoint via $\gc_0$ is no longer viable. Instead, we define $\bar{\Psi}= \Psi^{\dagger}\Lambda$, where $\Lambda$ is a Hermitian matrix to be determined by symmetry requirements. 

Under a Carroll transformation, the spinor $\Psi(x)$ transforms as 
	\begin{eqnarray}
		\Psi(x) \rightarrow \mathcal{S}[\Sigma]\Psi(x),
	\end{eqnarray}
where $\mathcal{S}[\Sigma] = \exp\big(\frac{1}{2}\omega \cdot \Sigma \big).$
Here $\Sigma$'s\footnote{When constructing either the lower or upper fermion theory, the indices of the Carroll generators $\Sigma$ are taken to be correspondingly lower or upper.} are the generators of the Carroll algebra defined via commutators of Carrollian gamma matrices and $\omega \equiv \omega^{\mu\nu}$'s are antisymmetric transformation parameters. Requiring Carroll invariance of the bilinear $\bar{\Psi}\Psi$ leads to the condition 
	\begin{equation}\label{sd}
		\Sigma^\dagger\Lambda + \Lambda\Sigma = 0.
	\end{equation}		
It can be checked that $(\ref{sd})$ is satisfied, provided $\Lambda$ obeys
\begin{subequations}
    \begin{eqnarray}\label{pm}
		&&\gc_{{\mu}}^\dagger = \pm \Lambda\gc_{\mu}\Lambda^{-1} \quad\text{(either + or -)},\\
		&&\Sigma_{\mu\nu}^\dagger = -\Lambda\Sigma_{\mu\nu}\Lambda^{-1}\label{pm1}.
	\end{eqnarray}
\end{subequations}
This is a non-trivial feature: for Lorentzian Dirac fermions, the adjoint matrix is embedded in the Clifford algebra itself, whereas in the Carrollian case it must be determined separately. A completely analogous set of relations also holds for $\gb$ and $\Sigma^{\mu\nu}$. 

\medskip

\subsection{Representations and actions}

Carroll Clifford algebras are intrinsically degenerate and contain nilpotent gamma matrices. Recall that an $n\times n$ matrix $A$ is nilpotent if $A^k =0$ for some positive integer $k$. By the rank-nullity theorem -- which states that for any linear map $T:V\to W$ between finite dimensional vector spaces, rank$(T)$ + nullity$(T)$ = dim$(V)$ -- the singularity of $A$ implies rank$(A)$ $\leq n-1$ \cite{Axler2024}. 

In two dimensions, this implies that the nilpotent $\gamma$'s have either rank $0$ or rank $1$, giving rise to two inequivalent representations. 
\begin{itemize}
    \item{\emph{Homogeneous representation:}} Of the two, the rank-$0$ realisation is referred to as the \textit{homogeneous} representation, denoted by $\mathcal{R}_H$. For $\mathcal{R}_H$, the nilpotent matrix is trivially zero, while the other gamma matrix spanning the Clifford algebra can be chosen among the Pauli spin matrices. In higher dimensions, $\mathcal{R}_H$ follows this same pattern. 
    \item{\emph{Inhomogeneous representation:}} On the other hand, the rank-$1$ realisation is called the \textit{inhomogeneous} representation, denoted by $\mathcal{R}_I$. Higher dimensional $\mathcal{R}_I$ can be systematically built from the lower dimensional gamma matrices. 
\end{itemize}

\paragraph{Lower gamma theory:} Let's first consider the lower gamma theory. The form of the Carrollian $\gc$ matrices in representation $\mathcal{R}_I$ in 2 and 4 spacetime dimensions as well as the adjoint matrix are as follows: 
\begin{subequations}
    \begin{align}
    \mathcal{R}_I \,(2D)\quad : \quad \gc_0 &= \begin{pmatrix}
        0 & 0\\
        1 & 0
    \end{pmatrix}\,,\quad \gc_1 = \begin{pmatrix}
        1 & 0\\
        0 & -1
    \end{pmatrix}\,,\quad \Lambda = \begin{pmatrix}
        0 & i\\
        -i & 0
    \end{pmatrix}\\
    \mathcal{R}_I \,(4D)\quad : \quad \gc_0 &= \begin{pmatrix}
        \textbf{0} & \textbf{0}\\
        \textbf{I} & \textbf{0}
    \end{pmatrix}\,,\quad \gc_i = \begin{pmatrix}
        \sigma^i & \textbf{0}\\
        \textbf{0} & -\sigma^i
    \end{pmatrix}\,,\quad \Lambda = \begin{pmatrix}
        \textbf{0} & i\textbf{I}\\
        -i\textbf{I} & \textbf{0}
    \end{pmatrix}
\end{align}
\end{subequations}
where $\textbf{I}$ is the $2\times 2$ identity matrix and $\sigma^i$ are the Pauli spin matrices \footnote{Note that we have chosen the rank-1 nilpotent gamma matrix in a lower-triangular form. One could alternatively select an upper-triangular form; however, it is important to ensure that both forms are not combined within the same algebra, as this would be inconsistent with the Clifford algebra.}. 
The corresponding massless fermion action, written in covariant form, is given by,
\begin{align}
    S_{\text{lower}} = \int d^Dx \,\bar{\Psi}\,\theta^{\mu}\theta^{\nu}\,\tilde{\gamma}_{\mu}\partial_{\nu}\, \Psi = \int dtd^{D-1}x \,\bar{\Psi}\gc_{0}\partial_{t}\Psi . 
\end{align}
For the homogeneous case, the action vanishes. In the inhomogeneous case $\gc_0$ being the rank 1 nilpotent matrix, only half of the spinor components contribute to the dynamics. This can be seen by writing $\Psi = \begin{pmatrix}
		\phi &~~~\chi
	\end{pmatrix}^T$
with $\phi$ and $\chi$ denoting equal sized blocks. The action reduces to
\begin{align}\label{lga}
    S_{\text{lower}} = \int dt d^{D-1}x \left(i \phi^\dagger \partial_{t}\phi \right).
\end{align}
Thus, in $D$-dimensions, a Carroll spinor carries half as many dynamical degrees of freedom compared to a relativistic spinor. Also as the action only contains time derivative and spatial derivatives vanish, effectively this behaves like a one-dimensional theory localized along the null direction.

\medskip

\paragraph{Upper gamma theory:} Now we look at the upper gamma theory. In 2 and 4 dimensions the $\gb$ matrices take the following form in $\mathcal{R}_I$ representation: 
\begin{subequations}
    \begin{align}
    \mathcal{R}_I \,(2D)\quad : \quad \gb^0 &= \begin{pmatrix}
        1 & 0\\
        0 & -1
    \end{pmatrix}\,,\quad \gb^1 = \begin{pmatrix}
        0 & 0\\
        1 & 0
    \end{pmatrix}\,,\quad \Lambda = \begin{pmatrix}
        0 & i\\
        -i & 0
    \end{pmatrix}\\
    \mathcal{R}_I \,(4D)\quad : \quad \gb^0 &= \begin{pmatrix}
        \textbf{I} & \textbf{0}\\
        \textbf{0} & -\textbf{I}
    \end{pmatrix}\,,\quad \gb^i = \begin{pmatrix}
        \textbf{0} & \textbf{0}\\
        \sigma^i & \textbf{0}
    \end{pmatrix}\,,\quad \Lambda = \begin{pmatrix}
        \textbf{0} & i\textbf{I}\\
        -i\textbf{I} & \textbf{0}
    \end{pmatrix}
\end{align}
\end{subequations}
In this case the action closely resembles the relativistic Dirac action and takes the form
\begin{align}
    S_{\text{upper}} = \int d^Dx \bar{\Psi}\gb^{\mu}\partial_{\mu}\Psi = \int dtd^{D-1}x\, \bar{\Psi}\left(\gb^0\partial_t + \gb^i\partial_i\right)\Psi\,.
\end{align}
Now, for the homogeneous case, only $\gb^0$ contributes in the action. Thus, we have a fermionic theory that retains both spinor components but only uses temporal derivatives. In the inhomogeneous scenario, both $\gb^0$ and $\gb^i$ take part in the dynamics and therefore both temporal and spatial derivatives come into play as can be seen by looking at the action 
\begin{subequations}\label{uga}
    \begin{align}
    &\text{In 2D:\qquad} S_{\text{upper}} = \int dtdx\,i\left(\phi^\star\dot{\chi} + \chi^\star\dot{\phi} - \phi^\star\partial_x{\phi}\right)\\
    &\text{In 4D:\qquad} S_{\text{upper}} = \int dtd^3x\,i\left(\phi^\dagger\dot{\chi} + \chi^\dagger\dot{\phi} - \phi^\dagger\sigma^i\partial_i{\phi}\right)\,.
\end{align}
\end{subequations}
We will show in next section that the two actions can be obtained as leading and subleading pieces of the relativistic fermionic action by appropriately expanding the spinor components. In modern parlance, these can be called electric and magnetic fermions as well.  

\subsection{Curiosities in odd dimensions}
So far, we have dealt with Carrollian gamma matrices in even spacetime dimensions. Let us now turn our attention to odd spacetime dimensions. Although the general covariant actions introduced above remains valid in any dimension, the dimensionality and structure of the representations of the Carroll Clifford algebra is intricate. 

\medskip

In relativistic theories, gamma matrices in odd dimensions $D=2N+1$ are typically constructed by defining an additional matrix as the product of all even-dimensional gamma matrices for $D=2N$ spacetime dimensions. This procedure yields a matrix that anti-commutes with all existing gamma matrices and squares to identity, thereby furnishing a consistent extension of the Clifford algebra without increasing the dimension of the representation. As a result, irreducible representations in odd dimensions are naturally inherited from those in one lower even spacetime dimensions. 

\medskip

In the Carrollian case, however, this construction is subtle and differs in lower and upper gamma representation. First consider the lower gamma inhomogeneous representation\footnote{In the homogeneous representation, certain gamma matrices vanish identically; consequently, their product with any other gamma matrix is zero by definition.}. Here, because of the presence of one non-zero nilpotent gamma matrix $\tilde{\g}_0$, the natural analogue of the relativistic construction, leads to a matrix
\begin{align}
    \tilde{\g}_5 = \prod_{a=0}^{d-1}\tilde{\g}_a.
\end{align}
whose square vanishes. Such a matrix cannot satisfy the Clifford algebra relation and therefore does not provide a consistent extension of the algebra to odd dimensions. This obstruction is a direct manifestation of the degeneracy of the Carrollian metric and has no analogue in relativistic Clifford algebra. 

In contrast, as in the upper gamma representation there exists only one gamma matrix, which squares to identity, a natural relativistic argument of defining 
\begin{align}
    \hat{\g}^5 = \prod_{a=0}^{d-1}\hat{\g}_a.
\end{align}
holds true here. Even though the square of it is still zero, this definition is consistent with the metric signature in the Clifford algebra. 

As a result, the classification of gamma matrices and spinor representations in odd Carrollian spacetime dimensions must be addressed independently, rather than naively following the relativistic argument. In particular, the presence of nilpotent gamma implies that the minimal dimension and the distinction between homogeneous and inhomogeneous representations require a separate and careful analysis. We provide details of this in Appendix \ref{appB}. We provide an example to close out this subsection and provide some context to what is to follow.

For concreteness, let's look at three spacetime dimensions. In this case, one finds that a 2d representation is insufficient to realise all Carroll Clifford relations consistently. A convenient way to proceed is to embed the 3D algebra into a 4D Clifford algebra and then remove one of the spatial gamma matrices. The remaining three gamma matrices then satisfy the appropriate Carroll Clifford relations in three spacetime dimensions. The price one pays is that the minimal representation is 4d rather than 2d, in sharp contrast with the relativistic case. It is as if Carroll fermions have flavours of a higher dimensional theory. As we will see later in Sec.~{\ref{sec4}} of this paper, this rather non-intuitive feature of Carroll fermions becomes very transparent when one studies this in lightcone coordinates. 

\subsection{Gamma-less Fermions} 
Finally, before moving onto Carroll expansions, we comment on a very different Carroll fermion, which to the best of our knowledge has not been mentioned anywhere in literature before. The simplest way to construct a Carroll invariant kinetic term for a spinor field is one that does not involve any gamma matrices at all, 
\begin{align}
    \mathcal{L} = i\bar{\Psi}\theta^{\mu}\partial_{\mu}\Psi.
\end{align}
As this Lagrangian is independent of Clifford algebra, its invariance is insensitive to the choice of the gamma matrix representation. The equation of motion for this is given by 
\begin{align}\label{gless}
   \theta^{\mu}\partial_{\mu}\Psi = 0. 
\end{align}
 Recall the action for a free massless electric Carroll scalar is
 \begin{align}
    S_{e} = \int d^dx dt \, (\partial_t \phi)^2. 
\end{align}
Its equations of motion are just 
\begin{align}\label{elec}
   \partial^2_t \phi  = 0. 
\end{align}
It is clear that the gamma-less fermion equation of motion \eqref{gless} upon squaring yields electric Carroll scalar equation \eqref{elec}. From this viewpoint, the spinor field behaves effectively as a Grassmann-valued Carroll scalar, with its spinorial character playing no dynamical role. This action may serve as a minimal Carrollian fermionic theory in contexts such as a building block for Carrollian supersymmetric theories. We leave a systematic investigation of its applications to future work.

\section{Carroll expansion of Dirac fermions}\label{sec3}
One of the characteristic features of a Carrollian structure is the collapse of the spacetime lightcone. In Lorentzian geometry, the slope of the lightcone is $1/c$, where $c$ denotes the speed of light; as $c\to 0$, the lightcone closes entirely and spatial propagation  In lieu of a contraction where one sends $c\to0$, Carrollian theories can be systematically derived through an expansion in small powers of $c$, allowing for a perturbative description around $c=0$ \cite{deBoer:2021jej, Hansen:2021fxi}. This is the natural Carroll counterpart of the post-Newtonian expansion, which organises corrections to Newtonian gravity in powers of $v^2/c^2 \ll 1$ and is behind many fundamental gravitational physics - for instance, the modelling of compact binary in-spiral, solar system ephemerides etc \cite{Blanchet_2002, Futamase:2007zz, Will_2014}. 

\medskip

As stated in the introduction, following this idea, the action of Carrollian scalars, spin one and spin two fields have been obtained from their relativistic counterparts. The basic building block is to assume that the relativistic bosonic field admits an expansion of the form 
\begin{align}\label{exp-bos}
\Phi(x) = c^\Delta \sum_{n=0}^\infty c^{2n} \phi^{(n)},
\end{align}
around $c=0$. Here $\Delta$ is an overall scaling associated with the field in question. This expansion is then plugged into the relativistic action and terms are collected order by order in powers of $c$. The leading term gives rise to the Lagrangian of the ``electric'' theory and this is manifestly Carroll invariant. The bare subleading term is a Lagrangian where Carroll boosts are broken. Carroll invariance is restored by adding Lagrange multipliers. We discuss this process in detail for scalars and the spin-one gauge field in the appendix \ref{appC}. We note in passing that the Carroll expansion of bosonic fields \eqref{exp-bos} was carried out for even powers of $c$. 

\medskip

We now demonstrate how Carrollian fermionic Lagrangians arise systematically from the relativistic Dirac theory through this controlled expansion in the power of $c$. Our analysis shows that different Carroll fermion theories emerge depending on how the relativistic spinor components are scaled in the Carroll limit. We begin with the massless Dirac Lagrangian given by,  
\begin{eqnarray}\label{Diraceqn}
\mathcal{L}_{rel} = \bar{\Psi}\gamma^{\mu}\partial_{\mu}\Psi
\end{eqnarray}
where the gamma matrices satisfy the relativistic Clifford algebra $\{\g^\mu,\g^\nu\} = 2\eta^{\mu\nu}$ and the Dirac adjoint $\bar{\Psi}$ is defined as $\bar{\Psi}=\Psi^\dagger i\g^0$. We work in 4D Minkowski spacetime with metric $\eta^{\mu\nu} = \text{diag}(-1,1,1,1)$ and coordinate system $x^{\mu} = (x^0,\vec{x}) = (ct,\vec{x})$. To express the action in terms of explicit spinor components, we choose a representation (say Weyl) of the $\gamma$-matrices and decompose the Dirac spinor into two two-component spinors $\Psi = \begin{pmatrix}
		\phi &~~~\chi
	\end{pmatrix}^T$ (for details, see appendix \ref{appA}). Under infinitesimal Lorentz transformations, the spinor $\Psi$ transforms as
\begin{align}
    \delta \Psi = -\omega^{\mu}_{\,\,\nu}x^{\nu}\partial_{\mu}\Psi + \frac{1}{2}\omega_{\rho\sigma}\Sigma^{\rho\sigma}\Psi\, ,
\end{align}
with the generators of the Lorentz transformations are  given by $\Sigma^{\rho\sigma}$ is defined as $\Sigma^{\rho\sigma} = \frac{1}{4}[\g^\rho,\g^\sigma]$.
The Lagrangian becomes 
\begin{equation}\label{DrCh}
	\mathcal{L}_{Weyl} = -\frac{1}{c}i(\phi^\dagger\dot{\phi} + \chi^\dagger\dot{\chi}) + i\phi^\dagger\s^i\partial_i\phi - i\chi^\dagger\s^i\partial_i\chi.
\end{equation}
We will be primarily interested in the action of boosts after we implement the Carroll expansion. For this, we note that in the relativistic fermions, the infinitesimal boost transformation of the spinor components are given by
\begin{subequations}
    \begin{align}\label{Chboo}
    &\delta_{LB} \phi = -ct\vec{\beta}\cdot\vec{\partial}\phi - \frac{1}{c}\vec{\beta}\cdot \vec{x}\partial_t \phi - \frac{1}{2}\vec{\beta}\cdot \vec{\sigma} \phi \, ,\\
    &\delta_{LB} \chi = -ct\vec{\beta}\cdot\vec{\partial}\chi - \frac{1}{c}\vec{\beta}\cdot \vec{x}\partial_t \chi + \frac{1}{2}\vec{\beta}\cdot \vec{\sigma} \chi\,.
\end{align}
\end{subequations}
In the above, $LB$ stands for Lorentz boosts. 

\subsection{Uneven expansion of spinor components}
In the Carroll expansion of the spin-one gauge field theory, which we discuss in Appendix~\ref{appC}, the temporal and spatial components of the gauge field are scaled differently. Motivated by this structure, it seems that natural to consider similar uneven expansions for fields with non-trivial spins. 

\medskip

For the Carroll expansions of spin-1/2 fermions, we thus first consider an uneven $c-$expansion of the two-component spinors around $c=0$:
\begin{align}\label{uneven}
    \phi = c^\Delta\sum_{n\geq 0}c^n\phi_n\,,\quad \chi = c^\Delta\sum_{n\geq 0}c^{n+1}\chi_n\,.
\end{align}
Here we introduce a relative factor of $c$ between two chiral components and $\Delta$ is some overall factor. Notice, that crucially, we don't assume an expansion in only even powers of $c$. Fermions necessitate odd powers. 

Substituting the ansatz in \eqref{DrCh}, and organizing terms order by order in powers of $c$, the Lagrangian takes the following form $\mathcal{L} = c^{\delta}\sum_nc^n\mathcal{L}_{(n)}$ with \footnote{Here we have absorbed an absolute minus sign prefactor in the left hand side.}
\begin{subequations}\label{car-lag}
    \begin{align}
    &\mathcal{O}(c^0): \quad \mathcal{L}_{(0)} = i\phi_0^\dagger\dot{\phi}_0 \label{LO}\\
    &\mathcal{O}(c^1): \quad \mathcal{L}_{(1)} = i\left(\phi_0^\dagger\dot{\phi_1} + \phi_1^\dagger\dot{\phi_0} - \phi_0^\dagger\s^i\partial_i\phi_0\right) \label{NLO}\\
    &\mathcal{O}(c^2): \quad \mathcal{L}_{(2)} = i\left(\phi_0^\dagger\dot{\phi_2} + \phi_2^\dagger\dot{\phi_0} + \phi_1^\dagger\dot{\phi_1} + \chi_0^\dagger\dot{\chi_0}\right) - i\left(\phi_0^\dagger\s^i\partial_i\phi_1 +\phi_1^\dagger\s^i\partial_i\phi_0\right)\,.
\end{align}
\end{subequations}
We now focus on the transformation under boosts. Defining the Carroll boost parameter $\vec{b}$ through $\vec{\beta} = c\vec{b}$, the boost transformations of the expanded spinor components are given by, 
\begin{subequations}\label{car-boost}
    \begin{align}
    &\mathcal{O}(c^0): \quad\delta_C \phi_0 = -\vec{b}\cdot\vec{x}\partial_t\phi_0\,,~~~~~~~~~~~~~~~~~~~~~~~~~~~~~~ \delta_C \chi_0 = -\vec{b}\cdot\vec{x}\partial_t\chi_0\label{lobt}\\
    &\mathcal{O}(c^1): \quad \delta_C \phi_1 = -\vec{b}\cdot\vec{x}\partial_t\phi_1 - \frac{1}{2}\vec{b}\cdot \vec{\sigma}\phi_0\,,~~~~~~~~~~~~~~~~\delta_C \chi_1 = -\vec{b}\cdot\vec{x}\partial_t\chi_1 + \frac{1}{2}\vec{b}\cdot \vec{\sigma}\chi_0\label{nlobt}\\
    &\mathcal{O}(c^2): \quad \delta_C \phi_2 = -\vec{b}\cdot\vec{x}\partial_t\phi_2 - t\vec{b}\cdot\vec{\partial}\phi_0 - \frac{1}{2}\vec{b}\cdot \vec{\sigma}\phi_1\,,\quad \delta_C \chi_2 = -\vec{b}\cdot\vec{x}\partial_t\chi_2 - t\vec{b}\cdot\vec{\partial}\chi_0 + \frac{1}{2}\vec{b}\cdot \vec{\sigma}\chi_1
\end{align}
\end{subequations}
Here $\delta_C$ represents the change of the fields under Carroll boosts. 

\medskip

By plugging in \eqref{car-boost} into the Lagrangian \eqref{car-lag}, it can be explicitly verified that \emph{both the leading order (LO) and the next-to leading order (NLO) Lagrangians} are invariant under these Carroll boosts (for details, refer to Appendix \ref{appD}). This persistence of Carroll invariance at sub-leading order is non-trivial and unlike the case of bosonic fields. As we stated above, in general, when performing a small $c$-expansion of a relativistic field theory, the LO Lagrangian is always Carroll boost invariant, but the sub-leading pieces are not automatically so. This is because the $c$-expansion on Lorentz boost transformation of fields typically mixes different orders in $c$, and so only the lowest consistent truncation is guaranteed to respect the Carroll symmetry. 

\medskip

In the fermionic case, however something special occurs: not only the LO theory, but the NLO theory also retains Carroll boost invariance. The reason of this can be traced back to the structure of the boost transformations. In $c$-expansion of the relativistic scalar (see Appendix \ref{appC}), the spatial derivative terms already show up at the sub-leading order in the boost transformations. This implies as soon as one goes beyond leading order, spatial dependence kicks in, and the sub-leading actions usually fail to be invariant under Carroll boosts. By contrast, for fermions, both the LO and NLO transformations involve only time derivatives and internal spin-rotations terms. The first explicit spatial dependence appears at only sub-subleading order which delays the breakdown of Carroll boost invariance. At LO, we see that the boost transformation of the spinor components only include the spatial boost term while the spin-rotation term enters only at NLO.

\medskip

\noindent The resulting equations of motion at successive orders are
\begin{subequations}
\begin{align}
    \mathcal{O}(c^0): &\quad \dot{\phi}_0 = 0 \\
    \mathcal{O}(c^1): &\quad \dot{\phi}_1 = \sigma^i\partial_i\phi_0\,,\quad  \dot{\phi}_0 = 0\\
    \mathcal{O}(c^2): &\quad \dot{\phi}_2 = \sigma^i\partial_i\phi_1\,,\quad \dot{\phi}_1 = \sigma^i\partial_i\phi_0\,,\quad  \dot{\phi}_0 = 0, \quad \dot{\chi}_0 = 0.
\end{align}    
\end{subequations}
As can be seen these equations of motion form a cascade like structure, where field in each order is sourced entirely by the spatial gradients of the field at the previous order. Appyling $\partial_t$ repeatedly
\begin{align}
    \ddot{\phi}_0 &= 0 \nonumber\\
    \ddot{\phi}_1 &= \sigma^i\partial_i\dot{\phi}_0 = 0\\
    \ddot{\phi}_2 &= \sigma^i\partial_i\dot{\phi}_1 = \sigma^i\partial_i\sigma^j\partial_j\phi_0 = \delta^{ij}\partial_i\partial_j\phi_0 = \nabla^2\phi_0\nonumber
\end{align}
we see at order $\mathcal{O}(c^n)$ the field $\phi_n$ is a degree-$n$ polynomial in $t$ built from successive applications of $\sigma^i\partial_i$ to $\phi_0$. Also at the order $\mathcal{O}(c^2)$,  $\chi_0$ appears and it's completely decoupled from the $\phi$ sector. It is a new independent electric Carroll field that becomes visible only at this order.

Carroll invariance of a field theory is directly reflected in its energy momentum tensor and a tell-tale feature is that the $(0i)$ component of the stress tensor vanishes. Below we will see this happening for the Carroll fermionic models we have encountered above. Computing energy-momentum tensor components using Noether's prescription
\begin{align}
    T^{\mu}_{\,\,\nu} = \frac{\p\mathcal{L}}{\p(\p_{\mu}\phi_a)}\p_{\nu}\phi_a - \delta^{\mu}_{\,\,\nu}\mathcal{L}\,,
\end{align}
we find that at different orders, the stress tensors are given by: 
\begin{subequations}
    \begin{align}
    \mathcal{O}(c^0): &\quad T^{0}_{\,\,0} = 0\,,\quad T^{i}_{\,\,0} = 0\,,\quad T^{0}_{\,\,i} = i\phi_0^\dagger\partial_i\phi_0\,,\quad T^{i}_{\,\,j} = -\delta^i_{\,\,j}\mathcal{L}\\
    \mathcal{O}(c^1): &\quad T^{0}_{\,\,0} = i\phi_0^\dagger\s^i\partial_i\phi_0\,,\quad \underbrace{T^{i}_{\,\,0} = -i\phi_0^\dagger\s^i\dot{\phi}_0}_{=0 \,\text{(onshell)}}\,,\quad T^{0}_{\,\,i} = i(\phi_1^\dagger\partial_i\phi_0 +\phi_0^\dagger\partial_i\phi_1), \nonumber \\
    &\quad T^{i}_{\,\,j} = -i\phi_0^\dagger\s^i\partial_j\phi_0 -\delta^i_{\,\,j}\mathcal{L}.
\end{align}
\end{subequations}
One finds that the component $T^{i}_{\,\,0}$ vanishes identically at leading order and at sub-leading order vanishes upon using equation of motion. This ensures the invariance of the theory under Carroll boosts.  

\medskip

\noindent\emph{Matching with intrinsic Carroll fermions:}  The leading \eqref{LO} and next-to-leading \eqref{NLO} Lagrangians precisely reproduce the structures of the intrinsic Carroll fermion actions constructed using the lower- and upper- gamma matrices, respectively. Hence we have 
\begin{align*}
&\boxed{\text{Lower gamma theory} = \text{ LO fermion in $c$-expansion (Electric theory)}} \\
&\boxed{\text{Upper gamma theory} = \text{NLO fermion in $c$-expansion (Magnetic theory)}}
\end{align*}
Moreover, the Carroll boost transformations obtained from the $c-$expansion \eqref{lobt} and \eqref{nlobt}, coincide with the intrinsic Carroll boost transformations acting on $\phi_0$ and $\phi_1$. This establishes a direct correspondence between the $c-$expansion and intrinsic Carrollian fermion dynamics.

We should comment here that comparing with the scalars and gauge fields, perhaps both LO and NLO theories could be considered ``Electric'', since they are invariant under Carroll boosts without having to invoke any Lagrange multipliers. The NNLO term yields 
\begin{align}
    \delta_C\mathcal{L}_{(2)} = -\partial_t(\vec{b}\cdot \vec{x}\mathcal{L}_{(2)}) - t\vec{b}\cdot\vec{\partial}(\phi_0^\dagger\dot{\phi}_0) \,.
\end{align}
This clearly breaks Carroll boosts and hence is more in the spirit of the ``magnetic'' theories that the scalar and gauge fields yield in their subleading expansions. It would be interesting to do a Hamiltonian analysis in the spirit of \cite{Henneaux:2021yzg} and understand Carroll fermions from the point of view of phase space. We hope to return to this in the near future. 

\subsection{Even expansion of spinor components}
Although, in keeping with the unequal expansions of the gauge fields, it seems natural to consider unequal expansions for all Carrollian fields with spin, we indulge ourselves in a bit of exploration.
We consider a symmetric expansion of the two-component spinors symmetrically around $c=0$
\begin{align}\label{even}
    \phi = c^\Delta\sum_{n\geq 0}c^n\phi_n\,,\quad \chi = c^\Delta\sum_{n\geq 0}c^{n}\chi_n\,.
\end{align}
Substituting this ansatz into \eqref{DrCh}, and expanding order by order in $c$ we obtain, 
\begin{subequations}
    \begin{align}
    &\mathcal{O}(c^0): \quad \mathcal{L}_{(0)} = i\left(\phi_0^\dagger\dot{\phi}_0\right) + i\left(\chi_0^\dagger\dot{\chi_0}\right) \label{LO1}\\
    &\mathcal{O}(c^1): \quad \mathcal{L}_{(1)} = i\left(\phi_0^\dagger\dot{\phi_1} + \phi_1^\dagger\dot{\phi_0} - \phi_0^\dagger\s^i\partial_i\phi_0\right) + i\left(\chi_0^\dagger\dot{\chi_1} + \chi_1^\dagger\dot{\chi_0} + \chi_0^\dagger\s^i\partial_i\chi_0\right)\label{NLO1}\\
    &\mathcal{O}(c^2): \quad \mathcal{L}_{(2)} = i\left(\phi_0^\dagger\dot{\phi_2} + \phi_2^\dagger\dot{\phi_0} + \phi_1^\dagger\dot{\phi_1}-\phi_0^\dagger\s^i\partial_i\phi_1 -\phi_1^\dagger\s^i\partial_i\phi_0\right)\nonumber\\ &~~~~~~~~~~~~~~~~~~~~~~~~~~~~~~~~~~+  i\left(\chi_0^\dagger\dot{\chi_2} + \chi_2^\dagger\dot{\chi_0} + \chi_1^\dagger\dot{\chi_1}+\chi_0^\dagger\s^i\partial_i\chi_1 +\chi_1^\dagger\s^i\partial_i\chi_0\right).
\end{align}
\end{subequations}
Several observations are immediate:
\begin{itemize}
    \item The leading order Lagrangian is Carroll boost invariant as expected. Additionally, \eqref{LO1} may be interpreted in two ways; either as two decoupled copies of lower gamma Carroll fermion action \eqref{lga} or as the upper gamma Carroll theory in the homogeneous representation for a suitable choice of $\gb^0$. 
    \item The next-to-leading order Lagrangian corresponds to two copies of the upper gamma Carroll fermion theory in the inhomogeneous representation. As in the uneven expansion, Carroll boost invariance persists at this order. 
\end{itemize}
The Carroll boost transformations can be derived order by order in $c$. Since the boosts act independently on the two spinor components (as we are in Weyl represenation for parent theory), their structures remain the same with those obtained in the uneven expansion, namely \eqref{lobt},\eqref{nlobt}. The agreement with the homogeneous upper-gamma representation follows from the absence of spin-rotation terms, which vanish identically in this case and render the transformation structure trivial. 

\medskip

\noindent The resulting equations of motion at successive orders are
\begin{subequations}
    \begin{align}
    \mathcal{O}(c^0): &\quad \dot{\phi}_0 = 0\,,\quad \dot{\chi}_0 = 0 \\
    \mathcal{O}(c^1): &\quad \dot{\phi}_1 = \sigma^i\partial_i\phi_0\,,\quad  \dot{\chi}_1 = -\sigma^i\partial_i\chi_0\,,\quad \dot{\phi}_0 =\dot{\chi}_0= 0\\
    \mathcal{O}(c^2): &\quad \dot{\phi}_2 = \sigma^i\partial_i\phi_1\,,\quad \dot{\phi}_1 = \sigma^i\partial_i\phi_0\,,\quad  \dot{\phi}_0 = 0 \nonumber\\
    &\quad \dot{\chi}_2 = -\sigma^i\partial_i\chi_1\,,\quad \dot{\chi}_1 = -\sigma^i\partial_i\chi_0\,,\quad  \dot{\chi}_0 = 0\,.
\end{align}
\end{subequations}

\noindent As before, we construct the energy momentum-tensor of the first two entries of this series of Lagrangians. Using Noether's prescription we get
\begin{subequations}
    \begin{align}
    \mathcal{O}(c^0): &\quad T^{0}_{\,\,0} = 0\,,\quad T^{i}_{\,\,0} = 0\,,\quad T^{0}_{\,\,i} = i\left(\phi_0^\dagger\partial_i\phi_0 + \chi_0^\dagger\partial_i\chi_0\right)\,,\quad T^{i}_{\,\,j} = -\delta^i_{\,\,j}\mathcal{L}\\
    \mathcal{O}(c^1): &\quad T^{0}_{\,\,0} = i\left(\phi_0^\dagger\s^i\partial_i\phi_0 - \chi_0^\dagger\s^i\partial_i\chi_0\right)\,,\quad \underbrace{T^{i}_{\,\,0} = -i\left(\phi_0^\dagger\s^i\dot{\phi}_0 -  \chi_0^\dagger\s^i\dot{\chi}_0\right)}_{=0 \,\text{(onshell)}}\,,\quad \nonumber \\ &\quad T^{0}_{\,\,i} = i\left(\phi_1^\dagger\partial_i\phi_0 +\phi_0^\dagger\partial_i\phi_1 + \chi_1^\dagger\partial_i\chi_0 +\chi_0^\dagger\partial_i\chi_1\right)\,, \nonumber \\
    & \quad T^{i}_{\,\,j} = -i\phi_0^\dagger\s^i\partial_j\phi_0 + i\chi_0^\dagger\s^i\partial_j\chi_0 -\delta^i_{\,\,j}\mathcal{L}.
\end{align}
\end{subequations}
We again note that at $\mathcal{O}(c^0)$, both the component $T^{0}_{\,\,0}$ and $T^{i}_{\,\,0}$ vanish identically. As mentioned earlier, the vanishing of the latter is a hallmark of Carroll boost invariance. The non-zero energy density $T^{0}_{\,\,0}$ appears at $\mathcal{O}(c^1)$, while the $T^{i}_{\,\,0}$ component vanishes in this case by virtue of equations of motion.

\subsection{Discussions}

From the above analysis, we observe that the relativistic fermionic theory can be systematically expanded in powers of $c$—either equally or unequally—between the two spinor components. The leading-order (LO) term in these expansions yields a theory invariant under Carrollian symmetry, corresponding to the \textit{electric} Carroll limit.
In the equal-power expansion, the LO Lagrangian corresponds to the upper Carroll fermion theory in the homogeneous representation. In this case, only the temporal gamma matrix survives, leading to an action that contains solely time derivatives of the spinor components. Since all spatial gamma matrices vanish identically, the associated boost and rotation generators also become trivial. Still, it remains a consistent and valid realization.
Effectively, with a single non-vanishing gamma matrix, one may choose it to coincide with the relativistic temporal gamma matrix without affecting either the Clifford algebra or the underlying Carroll algebra. This feature is further reflected in the transformation properties of the spinor components, where the LO expansion contains no spin-rotation terms.

Interestingly, there exists an alternative interpretation of the leading-order expansion, now in terms of the lower Carroll fermion theory. As discussed earlier, within the inhomogeneous representation, the nilpotent lower gamma matrix can be expressed in a Jordan block form. From this viewpoint, the LO Lagrangian can be regarded as describing two independent, non-overlapping copies of the lower Carroll fermion theory. This interpretation is also consistent with the transformation properties of the spinor components, even though in this case the boost and rotation generators remain non-vanishing.

In the unequal-power expansion, the LO theory retains only a single spinor component. This structure naturally corresponds to the lower Carroll fermion theory, where only one sector remains dynamical.

To gain some intuition about the structure, let us briefly examine the lower Carroll fermion theory itself. Here, the boost generators are non-zero, as we are working within the inhomogeneous representation where $\gb_0 \neq 0$.
The action of the Carroll boost on the spinor then takes the schematic form
\begin{align}
    C_i\begin{pmatrix}
        \psi_L\\
        \psi_R
    \end{pmatrix} = \begin{pmatrix}
        1 & 0\\
        \xi_i & 1
    \end{pmatrix}\begin{pmatrix}
        \psi_L\\
        \psi_R
    \end{pmatrix}
\end{align}
This shows that the $\psi_L$ component remains unchanged under Carroll boosts, while $\psi_R$ component mixes with $\psi_L$. 

\paragraph{Connecting with existing literature:} The authors in \cite{Bergshoeff:2023vfd} undertook a study of Carroll fermions, starting from relativistic Dirac fermions implementing the Carroll limit via carefully chosen field rescalings and projections, thereby obtaining two distinct classes of Carroll fermion theories. It is instructive to compare their results with our construction.
In their electric limit, the spinor transformations under Carroll boosts contain no spin-rotation term. This is consistent with our LO spinor transformation ($\delta_C\phi_0$), which likewise transforms without any spin-rotation contribution - a feature that originates directly from the nilpotency of the Carroll boost generator at the LO. Moreover, the resulting action retains only the time derivative sector governed by the relativistic $\gamma^0$. In upcoming work \cite{ABJM}, we explicitly show how this action arises naturally from the Carroll-Dirac gamma matrices. 
In their magnetic limit,  one of the spinor components acquires a spin-rotation term under Carroll boost, while the other transforms identically to the electric case. This is in direct correspondence with our sub-leading order transformations - $\delta_C\phi_1$ inherits a spin-rotation contribution while $\delta_C\phi_0$ remains unchanged. However, a precise identification at the level of action is more subtle: the structure of magnetic action in \cite{Bergshoeff:2023vfd} does not straightforwadrly match the action obtained from our construction, a careful reconciliation of the two approaches in the magnetic sector remains a open question that we defer to future work. 

\section{Light-cone fermions}\label{sec4}

In the previous section, we have given a detailed account of how the intrinsic Carrollian fermions, both the lower and the upper gamma theories, are connected to relativistic fermions in the same number of dimensions, by performing a systematic expansion in the speed of light of the relativistic Dirac theory. The connection of Carroll fermions to their relativistic counterparts living in the same number of dimensions was broadly along expected lines, although our expansion procedure uncovered some interesting and unconventional properties of Carroll fermions vis-a-vis bosonic Carrollian fields.  In the preceding sections, we also saw some curious features of Carrollian fermions in odd dimensions which seemed to indicate that these fermions have properties connected to relativistic fermions, now living in one higher dimension. Our goal in the current section is to establish how these properties manifest themselves by considering relativistic theories on the lightcone.  

In an effort to do so, we will investigate fermionic representations in light-cone coordinates and their relation to Carrollian symmetry. But before doing that, let's give a brief idea how Carrollian symmetry is encoded in the null realisation of the Poincaré algebra. 

\subsection{Poincare algebra in the lightcone}
The lightcone coordinates throw up many surprises. We begin by defining them: 
\begin{eqnarray}
    x^{\pm} = \frac{1}{\sqrt{2}}(t\pm z)\,,\qquad x^i=x^i\,,
\qquad
\partial_\pm=\frac{1}{\sqrt2}(\partial_t\pm\partial_z).
\end{eqnarray}
The Minkowski line element in these coordinates reads
 \begin{eqnarray}
    ds^2 = -dt^2 + \delta_{ab}\,dx^adx^b = -2dx^+dx^- + \delta_{ij}\,dx^idx^j, 
\end{eqnarray} 
where $(a,b = 1,2,3)$ and $(i,j = 1,2)$. We will now express the Poincare algebra in these coordinates. The Poincare generators in the light-cone basis are:
\begin{subequations}\label{poi-lc}
\begin{eqnarray}
&& P_- = \frac{1}{\sqrt{2}}(P_t - P_z )= \p_-, \quad P_+ =\frac{1}{\sqrt{2}}( P_t + P_z) = \p_+, \quad P_i = \p_i\\
&& J_{i-} =\frac{1}{\sqrt{2}}(J_{it} - J_{iz}) =  x^+\p_i + x_i\p_-, \quad J_{i+} = \frac{1}{\sqrt{2}}(J_{it} + J_{iz})  =  x^-\p_i + x_i\p_+, \\
&& J_{ij} = x_i\p_j - x_j\p_i, \qquad B= J_{zt} = x^+\p_+ - x^-\p_-.
\end{eqnarray}
\end{subequations}
We first take a historic detour. It was shown, famously by Susskind in the 1970's \cite{Susskind:1967rg}, that this lightcone 4D Poincare algebra contained in it a Galilean subalgebra: 
\begin{align}
\text{Gal}(1): ~\{P_+, J_{i-}, P_i, J_{ij} \}
\end{align}
This could be further enhanced to a Bargmann algebra which is the Galilean algebra with a mass extension so that: 
\begin{align}
[\text{Galilean Boost}_i, \text{Momentum}_j] = \delta_{ij} \times \text{Mass}.
\end{align}
where Mass is a central element of the algebra. In the case of the lightcone, $P_-$ acts as the mass parameter. This discovery was the key to the physics of the infinite momentum frame and the subsequent understanding of Discrete Lightcone Quantization leading to the BFSS conjecture and M-theory \cite{Banks:1996vh}. Note that there is another 3D Galilean subalgebra lurking in the 4D Poincare: 
\begin{align}
\text{Gal}(2): ~\{P_-, J_{i+}, P_i, J_{ij} \}
\end{align}
which can again be given a Bargmann lift by considering $P_+$ now as the mass. 
\begin{table}[h]
\centering
\renewcommand{\arraystretch}{2.2}
\resizebox{\textwidth}{!}{%
\begin{tabular}{|c||c|c|c|c|c|c|c|}
\hline
$[\text{col},\text{row}]$ 
& \cellcolor{carr1}$P_+$ 
& \cellcolor{carr1}$J_{k+}$ 
& \cellcolor{carr12}$P_k$ 
& \cellcolor{carr12}$J_{kl}$ 
& \cellcolor{carr2}$J_{k-}$ 
& \cellcolor{carr2}$P_-$ 
& $B$ \\
\hline\hline
\cellcolor{carr1}$P_+$ 
& \cellcolor{carr1}$0$ 
& \cellcolor{carr1}$0$ 
& \cellcolor{carr1}${0}$ 
& \cellcolor{carr1}${0}$ 
& \cellcolor{white}$P_k$ 
& \cellcolor{white}$0$ 
& $P_+$ \\
\hline
\cellcolor{carr1}$J_{i+}$ 
& \cellcolor{carr1}$0$ 
& \cellcolor{carr1}$0$ 
& \cellcolor{carr1}${-\delta_{ik}P_+}$ 
& \cellcolor{carr1}${\delta_{ik}J_{l+} - \delta_{il}J_{k+}}$ 
& \cellcolor{white}$-\delta_{ik}B - J_{ik}$ 
& \cellcolor{white}$-P_i$ 
& $J_{i+}$ \\
\hline
\cellcolor{carr12}$P_i$ 
& \cellcolor{carr1}${0}$ 
& \cellcolor{carr1}${\delta_{ik}P_+}$ 
& \cellcolor{carr12}$0$ 
& \cellcolor{carr12}$\delta_{ik}P_l - \delta_{il}P_k$ 
& \cellcolor{carr2}${\delta_{ik}P_-}$ 
& \cellcolor{carr2}${0}$ 
& $0$ \\
\hline
\cellcolor{carr12}$J_{ij}$ 
& \cellcolor{carr1}${0}$ 
& \cellcolor{carr1}${\delta_{ik}J_{l+} - \delta_{jk}J_{l+}}$ 
& \cellcolor{carr12}$\delta_{jk}P_i - \delta_{ik}P_j$ 
& \cellcolor{carr12}$\begin{array}{c}\delta_{jk}J_{il} - \delta_{ik}J_{jl} \\ +\,\delta_{il}J_{jk} - \delta_{jl}J_{ik}\end{array}$ 
& \cellcolor{carr2}${\delta_{il}J_{k-} - \delta_{ik}J_{l-}}$ 
& \cellcolor{carr2}${0}$ 
& $0$ \\
\hline
\cellcolor{carr2}$J_{i-}$ 
& \cellcolor{white}$-P_i$ 
& \cellcolor{white}$\delta_{ik}B + J_{ik}$ 
& \cellcolor{carr2}${-\delta_{ik}P_-}$ 
& \cellcolor{carr2}${\delta_{ik}J_{l-} - \delta_{il}J_{k-}}$ 
& \cellcolor{carr2}${0}$ 
& \cellcolor{carr2}${0}$ 
& $-J_{i-}$ \\
\hline
\cellcolor{carr2}$P_-$ 
& \cellcolor{white}$0$ 
& \cellcolor{white}$-P_i$ 
& \cellcolor{carr2}${0}$ 
& \cellcolor{carr2}${0}$ 
& \cellcolor{carr2}$0$ 
& \cellcolor{carr2}$0$ 
& $-P_-$ \\
\hline
$B$ 
& $-P_+$ 
& $-J_{i+}$ 
& $0$ 
& $0$ 
& $J_{i-}$ 
& $P_-$ 
& $0$ \\
\hline
\end{tabular}%
}
\caption{Commutation relations $[\text{col},\text{row}]$ of light-cone generators. \\
\colorbox{carr1}{\strut Blue} and \colorbox{carr2}{\strut Green} blocks = {Carr}($\pm$) and \colorbox{carr12}{\strut Purple} block =shared subalgebra $\{P_i, J_{ij}\}$.}
\label{tab:commutators}
\end{table}

Interestingly, in recent work \cite{Bagchi:2024epw, Majumdar:2024rxg} it was found that the Poincare algebra not only contains two Galilean algebras, but also two 3D Carrollian subalgebras: 
\begin{subequations}
    \begin{align}
        &\text{Carr}(+):~\{P_+, P_i, J_{i+}, J_{ij}\},\\
        &\text{Carr}(-):~\{P_-, P_i, J_{i-}, J_{ij}\}.
    \end{align}
\end{subequations}
We highlight the Carrollian algebras in the table of commutators above. We see that the two subalgebras Carr$(\pm)$ correspond to two lightlike directions $x^\pm$, where, e.g. for the $+$ direction,  the Hamiltonian is $P_+$ and the Carroll boost is $J_{i+}$. However, these algebras have cross commutators between themselves, e.g. 
\begin{align}
[J_{i+}, J_{j-}] = - B\delta_{ij} - J_{ij}. 
\end{align}   
To focus on physics on the null surface $x^\pm$, we rescale one of the null coordinates, say $x^-\to\epsilon x^-$ and then take the limit $\epsilon\to0$ \footnote{We could have chosen $x^+$ instead. The results will be same.}. In this limit, the generators become
\begin{subequations}
    \begin{align}
        &H^{(-)}=\lim_{\epsilon\to0}\epsilon P_-=\partial_-,\quad H^{(+)}=\partial_+,\quad P_i=\partial_i,\\ &
        C^{(-)}_{i}=\lim_{\epsilon\to0} \epsilon J_{i-}=x_i\partial_-,\quad C^{(+)}_{i}=\lim_{\epsilon\to0}J_{i+}= x_i\partial_+,\\
        &J_{ij}= x_i\p_j - x_j\p_i,,\quad B=x^+\partial_+-x^-\partial_-.
    \end{align}
\end{subequations}
The identification of the Carroll sub-algebras remain the same 
\begin{subequations}\label{carrsuba}
    \begin{align}
        &\text{Carr}(+):~\{ H^{(+)}, P_i,C^{(+)}_{i}, J_{ij}\},\\
        &\text{Carr}(-):~\{H^{(-)}, P_i,C^{(-)}_{i}, J_{ij}\}.
    \end{align}
\end{subequations}
However, crucially now, the algebras decouple:
\begin{align}
[\mathcal{O}^{(+)}, \mathcal{O}^{(-)}] = 0
\end{align}
where $\mathcal{O}^{(\pm)} = H^{(\pm)}, C^{(\pm)}_{i}$. 
We also note the interesting role of the generator $B$. This survives the contraction and acts as similar to a dilatation operator with the following non-trivial commutation relations,
\begin{eqnarray}\label{B-act}
    [\mathcal{O}^{(\pm)}, B] = \pm \mathcal{O}^{(\pm)}, \quad [\mathcal{O}^{(T)}, B] = 0. 
\end{eqnarray}
where $\mathcal{O}^{(T)}= P_i, J_{ij}$ are the so-called ``transverse" generators, common to both algebras. 

Physically, the null contraction effectively suppresses dynamics along one light-cone direction while retaining a Carrollian structure on null hypersurfaces. This will be particularly relevant in our later discussion of fermions in light-cone backgrounds. 

\subsection{Light-cone representation of Clifford algebra}
We now focus on fermions and specifically on the Clifford algebra and its rewriting in the lightcone to understand whether similar lower dimensional non-Lorentzian structures arise. Working in 4D Minkowski spacetime, we introduce light-cone combinations of the gamma matrices with lower indices,
\begin{equation}
\gamma_{\pm} = \frac{1}{\sqrt{2}}(\gamma_0 \pm \gamma_3).
\end{equation}
These satisfy the algebra
\begin{equation}
\{\gamma_+,\gamma_-\} = -2\,, 
\qquad \{\gamma_\pm,\gamma_\pm\}=0\,,
\qquad \{\gamma_\pm,\gamma_i\}=0\,.
\end{equation}
Indices are raised and lowered using $\eta_{\mu\nu}$, so that $\gamma_\pm=-\gamma^{\mp}$.
The light-cone decomposition naturally leads to two independent sets of gamma matrices
\begin{subequations}
\begin{align}
(+):\; &\{\gamma_+,\gamma_1,\gamma_2\}, \\
(-):\; &\{\gamma_-,\gamma_1,\gamma_2\},
\end{align}
\end{subequations}
each of which satisfies a degenerate Clifford algebra
\begin{equation}\label{degclif}
\{\gamma_{\mu},\gamma_{\nu}\} = 2 \mathcal{G}_{\mu\nu},
\qquad
\mathcal{G}_{\mu\nu} = \mathrm{diag}(0,1,1).
\end{equation}
This degeneracy reflects the underlying Carrollian structure associated with the null directions in light-cone coordinates. We thus see that the lower dimensional Carrollian/Galilean subgroups that arise in the lightcone Poincare algebra also manifest themselves in terms of the relativistic Clifford algebra which too splits rather nicely into two lower dimensional Carrollian Clifford algebras.

\medskip

\noindent To see this even more explicitly, consider the corresponding spin generators defined as
\begin{align}\label{spingen}
S_{\pm i} &= \frac{1}{4}[\gamma_{\pm},\gamma_i], \qquad
J = \frac{1}{4}[\gamma_1,\gamma_2].
\end{align}
In the Weyl basis, the gamma matrices take the explicit form
\begin{align}
\gamma_- &= \sqrt{2}
\begin{pmatrix}
0 & -\sigma\\
\tilde{\sigma} & 0
\end{pmatrix}, \qquad
\gamma_+ = \sqrt{2}
\begin{pmatrix}
0 & -\tilde{\sigma}\\
\sigma & 0
\end{pmatrix}, \qquad
\gamma_i =
\begin{pmatrix}
0 & \sigma^i\\
\sigma^i & 0
\end{pmatrix}, \qquad i=1,2 ,
\end{align}
where we have defined 
\begin{equation}
\sigma^{\pm} = \frac{1}{2}(\sigma^1 \pm i\sigma^2), \qquad
\sigma = \frac{1}{2}(\sigma^0 + \sigma^3), \qquad
\tilde{\sigma} = \frac{1}{2}(\sigma^0 - \sigma^3).
\end{equation}
In this representation, the generators become
\begin{align}\label{Carr3d}
S_{-x} &= \frac{1}{\sqrt{2}}
\begin{pmatrix}
-\sigma^+ & 0\\
0 & \sigma^-
\end{pmatrix}, \qquad
S_{-y} = \frac{1}{\sqrt{2}}
\begin{pmatrix}
i\sigma^+ & 0\\
0 & i\sigma^-
\end{pmatrix}, \qquad
J = \frac{i}{2}
\begin{pmatrix}
\sigma^3 & 0\\
0 & \sigma^3
\end{pmatrix}\nonumber\\
S_{+x} &= \frac{1}{\sqrt{2}}
\begin{pmatrix}
-\sigma^- & 0\\
0 & \sigma^+
\end{pmatrix}, \qquad
S_{+y} = \frac{1}{\sqrt{2}}
\begin{pmatrix}
 -i\sigma^- & 0\\
0 &  -i\sigma^+
\end{pmatrix}.
\end{align}
These generators can equivalently be obtained from the light-cone decomposition of the Lorentz generators in the following way,
\begin{subequations}
    \begin{align}\label{Lor4d}
    &S_{+i} = \frac{1}{\sqrt{2}}\left(S_{ti} - S_{iz}\right)\,,\qquad S_{-i} = \frac{1}{\sqrt{2}}\left(S_{ti} + S_{iz}\right)\\
    & J = S_{xy}\,\qquad V = S_{tz} \qquad\qquad i = \{x,y\}
\end{align}
\end{subequations}
They obey the commutation relations
\begin{subequations}
    \begin{align}
    &[S_{+i},S_{+j}] = 0\,,\qquad [S_{-i},S_{-j}] = 0\\
    &[S_{\pm i},J] = \epsilon_{ij}S_{\pm j} \,, \qquad [S_{+i},S_{-j}] = J + \delta_{ij}V
\end{align}
\end{subequations}
with $\epsilon_{xy} = +1$. We thus see clearly that the Lorentz algebra thus contains two overlapping co-dimension one Carroll (or Galilei) subalgebras
\begin{align}
    & \mathcal{A}_+ : \{S_{+i}, J\}\,,\qquad
     \mathcal{A}_- : \{S_{-i}, J\}.
\end{align}
We should clarify here that the non-Lorentzian algebras are the versions where translation generators are absent. The commutators of the rotations with Carroll or Galilean boosts remain the same leading to isomorphic algebras. Hence the lower dimensional Clifford algebras correspond to both the 3D Carrollian and Galilean subalgebras of 4D Poincare algebra. 

\subsection{Spinor decomposition and transformations}
We now analyze how Dirac spinors transform under the Carrollian subalgebras identified above. Since the rotation symmetry is generated by a single spatial rotation $J$, its spinorial representation takes the block-diagonal form
\begin{eqnarray}
    S[J] = \begin{pmatrix}
        e^{i \varphi^z \sigma^3/2} & 0\\
        0 & e^{i \varphi^z \sigma^3/2}
    \end{pmatrix}
\end{eqnarray}
where $\varphi^z$ denotes the rotation parameter. The reducibility of this representation allows us to decompose the 4-component spinor $\Psi$ as
\begin{eqnarray}
    \Psi= \begin{pmatrix}
        \psi_+ \\
        \psi_-
    \end{pmatrix} = \begin{pmatrix}
        \psi_{+A}\\
        \psi_{+B}\\
        \psi_{-A}\\
        \psi_{-B}
    \end{pmatrix}\,.
\end{eqnarray}
Under spatial rotation, they transform identically, 
\begin{eqnarray}
    \psi_{\pm} \to e^{i \varphi^z \sigma^3/2} \psi_{\pm}\,,
\end{eqnarray}
reflecting the fact that the rotation generator is unaffected by the lightcone decomposition. In contrast, Carroll (or equivalently Galilean) boosts act differently on the two spinor components, depending on which co-dimension one subalgebra is chosen. 

\paragraph{Representation 1 (R1).} For the algebra generated by the set $\mathcal{A}_-$, the boost operators are represented as
\begin{eqnarray}\label{rep1}
    \mathcal{S}[S_{-x}] = \begin{pmatrix}
        e^{ \vartheta_x \sigma^+/\sqrt{2}} & 0\\
        0 & e^{ -\vartheta_x \sigma^-/\sqrt{2}}
    \end{pmatrix}\,,\qquad
    S[S_{-y}] = \begin{pmatrix}
        e^{- i\vartheta_y\sigma^+/\sqrt{2}} & 0\\
        0 & e^{ -i\vartheta_y\sigma^-/\sqrt{2}}
    \end{pmatrix}
\end{eqnarray}
Their act on the spinor components read,
\begin{subequations}\label{Carrb1}
  \begin{eqnarray}
    \text{along }x&&: \quad \psi_{+}^{(1)} \to e^{ \vartheta_x \sigma^+/\sqrt{2}} \psi_{+}^{(1)} \qquad \psi_{-}^{(1)} \to e^{-\vartheta_x \sigma^-/\sqrt{2}} \psi_{-}^{(1)}\\
    \text{along }y&&: \quad \psi_{+}^{(1)} \to e^{- i\vartheta_y\sigma^+/\sqrt{2}} \psi_{+}^{(1)} \qquad \psi_{-}^{(1)} \to e^{-i\vartheta_y\sigma^-/\sqrt{2}} \psi_{-}^{(1)}
\end{eqnarray}   
\end{subequations}
Here and in what follows, the superscript $(1)$ and $(2)$ label which representation, or more specifically which null surfaces, we are working on. Expanding to first order in the boost parameters $\vartheta_x$ and $\vartheta_y$, the transformations in four component notation become
\begin{subequations}
\begin{align}
\text{along }x:& \quad
\psi_{+A}\to\psi_{+A}+\frac{v_x}{\sqrt{2}}\psi_{+B}, \qquad
\psi_{+B}\to\psi_{+B}, \nonumber\\
& \quad
\psi_{-A}\to\psi_{-A}, \qquad
\psi_{-B}\to\psi_{-B}-\frac{v_x}{\sqrt{2}}\psi_{-A}.\\
\text{along }y:& \quad
\psi_{+A}\to\psi_{+A}-\frac{i v_y}{\sqrt{2}}\psi_{+B}, \qquad
\psi_{+B}\to\psi_{+B}, \nonumber\\
& \quad
\psi_{-A}\to\psi_{-A}, \qquad
\psi_{-B}\to\psi_{-B}-\frac{i v_y}{\sqrt{2}}\psi_{-A}.
\end{align}
\end{subequations}
We see that in R1, under these boost transformations, the $(1,4)$ components of the spinor transform and the $(2,3)$ components remain untouched. It is possible to say that in the four component notation, the $(1,4)$ components live of $x^-$ and $(2,3)$ live on $x^+$. The action of  $\mathcal{A}_-$ thus leaves the spinor components on $x^+$ untouched and only transform the components along $x^-$, as expected.

\medskip

\paragraph{Representation 2 (R2).} For the algebra generated by the set $\mathcal{A}_+$, the boost operators are represented as
\begin{eqnarray}\label{rep2}
    S[S_{+x}] = \begin{pmatrix}
        e^{ \vartheta_x \sigma^-/\sqrt{2}} & 0\\
        0 & e^{ -\vartheta_x \sigma^+/\sqrt{2}}
    \end{pmatrix}\,,\qquad
    S[S_{+y}] = \begin{pmatrix}
        e^{ i\vartheta_y\sigma^-/\sqrt{2}} & 0\\
        0 & e^{ i\vartheta_y\sigma^+/\sqrt{2}}
    \end{pmatrix}
\end{eqnarray}
Their act on the spinor components read,
\begin{eqnarray}\label{Carrb2}
    \textit{along x}&&: \quad \psi_{+}^{(2)} \to e^{ \vartheta_x \sigma^-/\sqrt{2}} \psi_{+}^{(2)} \qquad \psi_{-}^{(2)} \to e^{-\vartheta_x \sigma^+/\sqrt{2}} \psi_{-}^{(2)}\\
    \textit{along y}&&: \quad \psi_{+}^{(2)} \to e^{ i\vartheta_y\sigma^-/\sqrt{2}} \psi_{+}^{(2)} \qquad \psi_{-}^{(2)} \to e^{i\vartheta_y\sigma^+/\sqrt{2}} \psi_{-}^{(2)}
\end{eqnarray}
Expanding again to linear order, one finds
\begin{align}
\text{along }x:& \quad
\psi_{+A}\to\psi_{+A}, \qquad
\psi_{+B}\to\psi_{+B}+\frac{v_x}{\sqrt{2}}\psi_{+A}, \nonumber\\
& \quad
\psi_{-A}\to\psi_{-A}-\frac{v_x}{\sqrt{2}}\psi_{-B}, \qquad
\psi_{-B}\to\psi_{-B}.\\
\text{along }y:& \quad
\psi_{+A}\to\psi_{+A}, \qquad
\psi_{+B}\to\psi_{+B}+\frac{i v_y}{\sqrt{2}}\psi_{+A}, \nonumber\\
& \quad
\psi_{-A}\to\psi_{-A}+\frac{i v_y}{\sqrt{2}}\psi_{-B}, \qquad
\psi_{-B}\to\psi_{-B}.
\end{align}
In contrast to R1, R2 leaves the $(1,4)$ components of the spinor untouched and transforms the $(2,3)$ components. This is again consistent with the statement that the $(1,4)$ components live on $x^-$ and $(2,3)$ along $x^+$. R2, which corresponds to the set $\mathcal{A}_+$, thus only acts on the $(2,3)$ and not on $(1,4)$.  

\subsection{Null limit of Dirac action}
The null contraction that we described in the beginning of the section focusses the underlying theory on the null directions and decouples the two different Carroll subalgebras of the lightcone Poincare algebra. We have seen that the relativistic Clifford algebra naturally splits into its degenerate lower dimensional cousins without having to evoke the limit. However to understand fermionic field theories and focus on the null directions, it is again important to perform the null contraction. This is what we do in the current sub-section.

\medskip

We now derive the Carrollian limit of the massless Dirac theory in lightcone coordinates. The Dirac Lagrangian can be written as 
\begin{align}
    \mathcal{L} = i\bar{\Psi}\left(\gamma^+ \partial_+ + \gamma^- \partial_- + \gamma^i \partial_i\right)\Psi .
\end{align}
Now we perform the contraction
\begin{equation}
x^- \rightarrow \epsilon x^-, \qquad x^+ \rightarrow x^+, \qquad x^i \rightarrow x^i,
\end{equation}
and take $\epsilon \to 0$. Under this scaling, derivatives transform as $\partial_- \to \epsilon^{-1}\partial_-$, while $\partial_+$ and $\partial_i$ remain finite. The Lagrangian reduces to 
\begin{equation}\label{CarrL}
\mathcal{L} = i\bar{\Psi}\gamma^- \partial_- \Psi
= -i\sqrt{2}\psi_+^\dagger \sigma \partial_- \psi_+
  -i\sqrt{2}\psi_-^\dagger \tilde{\sigma} \partial_- \psi_- .
\end{equation}
The resulting equations of motion are 
\begin{eqnarray}\label{carrLeom}
    i\sigma\partial_-\psi_+ = 0\,,\qquad 
    i\tilde{\sigma}\partial_-\psi_- = 0
\end{eqnarray}
These describe decoupled Weyl spinors whose dynamics are localized along the $x^-$ null direction and constitute the fermionic realization of Carroll dynamics. 

\medskip

\noindent One may alternatively perform the opposite null contraction
\begin{eqnarray}
    x^+ \to \epsilon x^+\,,\quad x^- \to  x^-\,, \quad x^i \to x^i\,,
\end{eqnarray}
which leads, in the $\epsilon \to 0$ limit, to the Lagrangian
\begin{eqnarray}\label{L2}
    \mathcal{L} = i\bar{\Psi} \gamma^+ \partial_+ \Psi = -i\sqrt{2}\psi_-^\dagger\sigma\partial_+\psi_- -i\sqrt{2}\psi_+^\dagger\tilde{\sigma}\partial_+\psi_+
\end{eqnarray}
The corresponding equations of motion are 
\begin{eqnarray}\label{eqm2}
    i\sigma\partial_+\psi_- = 0\,,\qquad
    i\tilde{\sigma}\partial_+\psi_+ = 0\,.
\end{eqnarray}
This choice yields an equivalent Carrollian theory with the roles of two null directions swapped. 

\subsection{Symmetries of the theory}
We now investigate the symmetries of the null-contracted Dirac theory. We will carefully analyse the various symmetries associated with the action. We will discuss continuous spacetime symmetries, discrete symmetries and internal symmetries, as well as the action of the generator $B$ which is the $tz$-boost generator written in lightcone coordinates and acts similar to a dilatation generator.

\subsubsection{Spacetime symmetries}
We begin with spacetime symmetries of the Lagrangian \eqref{CarrL}. As is expected, this is invariant under the full 3D Carroll group. For completeness we note the transformations below:
\begin{itemize}
\item \textit{Translations}
\begin{equation}
x^- \to x^- + a^- , \qquad
x^i \to x^i + a^i ,
\end{equation}
since the action depends only on $\partial_-$.

\item \textit{Spatial rotations}
\begin{equation}
x^i \to R^i{}_j x^j , \qquad
\psi_\pm \to \mathcal{R}\,\psi_\pm ,
\end{equation}
where $\mathcal{R}$ denotes the spinor representation of $SO(D-2)$. These leave the null derivative $\partial_-$ invariant.

\item \textit{Carroll boosts}
\begin{equation}
    x^i \to x^i ,\qquad x^- \to x^- + v_ix^i ,\qquad  \delta\psi_{\pm} = -v_ix^i\partial_-\psi_{\pm} \pm\frac{1}{\sqrt{2}}(v_x \mp iv_y)\sigma^{\pm}\psi_{\pm}
\end{equation}
The above mentioned transformations show the invariance under the finite dimensional 3D Carrollian algebra, which is of course expected. 

\item \textit{Supertranslations}
\begin{equation}
x^i \to x^i  , \qquad
x^- \to x^-  + f(x^i)
\end{equation}
under which the Lagrangian remains invariant due to the absence of transverse derivatives. We thus get an infinite dimensional extension of symmetries with these supertranslations, which are angle dependent translations of the null direction $x^-$ \eqref{CarKill}. 

\end{itemize}

The null Dirac action does not have a mass term associated with it and hence it is plausible that there is an enhancement to conformal Carrollian symmetries in this context. We first write down the generators of these transformations (on the null surface):
\begin{align}
    D&= x^-\partial_- + x^i\partial_i \,, \quad K_-= x^kx_k\partial_- \,, \quad K_i = 2x_i (x^-\partial_- + x^k\partial_k) - x^kx_k\,\partial_i. 
\end{align}
We now check for the symmetry of the action under these generators:
\begin{itemize}

\item \textit{Dilatation}
\begin{equation}\label{scaling}
    x^- \to x^- + \lambda x^-, \qquad 
    x^i \to x^i + \lambda x^i, \qquad 
    \delta\psi_\pm = \lambda\left(x^-\partial_- + x^i\partial_i + \Delta\right)\psi_\pm,
\end{equation}
where $\lambda$ denotes the parameter for the scaling transformation and $\Delta$ is the conformal weight of the spinor field, which is found to be equal to 1, upon demanding the invariance of the action.  

\item \textit{Special conformal transformation $K_-$}
\begin{equation}
    x^- \to x^- + b\,x^ix_i, \qquad 
    x^i \to x^i, \qquad
    \delta_{K_-} \psi_\pm = -bx^ix_i\partial_-\psi_{\pm}  \pm b\sqrt{2}(\mp x + iy)\sigma^{\pm}\psi_{\pm}
\end{equation}
The variation gives the total derivative $\delta_{K_-}\mathcal{L} = \partial_{-}(-bx^ix_i\mathcal{L})$. 

\item \textit{Special conformal transformations $K_i$}
\begin{align}
    &x^- \to x^- + 2b^i x_i x^-, \qquad 
    x^j \to x^j + b^i(2x_ix^j - x^kx_k\,\delta^j{}_i),\nonumber\\
    &\delta_{K_i}\psi_\pm = b^i\left(2x_i (x^-\partial_- + x^k\partial_k) - x^kx_k\,\partial_i + 2\Delta x_i\right)\psi_\pm \nonumber \\ & \qquad \qquad -\sqrt{2}x^-(\mp b^x+ib^y)\sigma^{\pm}\psi_{\pm} + i\epsilon_{ij}b^ix^j\sigma^3\psi_{\pm}
\end{align}
The variation gives 
\begin{align}
    \delta\mathcal{L} = \partial_-\left(2b^ix_i x^-\mathcal{L}\right) + \partial_k\left(2b^ix_i x^k\mathcal{L}\right)
    - \partial_i\left(b^ix^kx_k\,\mathcal{L}\right) + 4b^i(1-\Delta)x^i\mathcal{L}
\end{align}
which vanishes provided $\Delta=1$. 
\end{itemize}
The Lagrangian \eqref{L2} is thus invariant under the finite dimensional Conformal Carroll transformations. 

The conformal Carroll algebra is the Lie algebra corresponding to conformal isometries of flat Carroll manifold
\begin{align}
    \pounds_\xi \, h_{\mu\nu} = \lambda_1 h_{\mu\nu}\,,\quad \pounds_\xi \,\theta^\mu = \lambda_2 \theta^\mu
\end{align}
where $\lambda_i$'s are constants. This is naturally infinite dimensional with the supertranslations that we encountered earlier  
(we take $\lambda_1/\lambda_2 = -2$): 
\begin{align}
    \xi  = f(x)\p_t + \omega^i_{\,\,j}x^j\p_i + b^i\p_i + \Delta\left(x^i\p_i + t\p_t \right) + k_i\left(2x^i\left(x^k\p_k + t\p_t \right) - x^kx_k\delta^{ij}\p_j\right).
\end{align}
Here $\omega^i_{\,\,j}, b^i, \Delta, k_i$ are integration of constants. Lie algebra of these vector fields is also famously the Bondi-van der Burg-Metzner-Sachs (BMS) algebra \cite{Bondi:1962px, Sachs:1962wk} which arises as asymptotic symmetries of asymptotically flat spacetimes in $(D+1)$ dimensions, giving an isomorphism between the Conformal Carroll and BMS algebras \cite{Bagchi:2010eg, Duval:2014uva}:
\begin{align}
    \mathfrak{CCarr}_D = \mathfrak{bms}_{D+1}.
\end{align}
This algebra is realised on the null boundary of asymptotically flat spacetimes, which has the topology of a cylinder $\mathbb{R} \times \mathbb{S}^{D-1}$, where $\mathbb{R}$ is the null direction. In $D=3$, there is an additional enhancement of the conformal symmetry to two copies of the Virasoro algebra on the 2-sphere. The generators of this doubly infinite extended algebra, sometimes called the extended BMS or $\mathfrak{ebms}_4$ \cite{Barnich:2009se, Barnich:2010eb} reads
\begin{align}
    L_n = z^{n+1}\partial_z + \frac{1}{2}(n+1)z^nt\partial_t\,, \quad \bar{L}_n =\bar{z}^{n+1}\partial_{\bar{z}} + \frac{1}{2}(n+1)\bar{z}^n t\partial_t \,, \quad M_{r,s} = z^r\bar{z}^s\partial_t
\end{align}
These generators close onto the $\mathfrak{ebms}_4$ algebra, the non-zero commutators of which are given by:
\begin{subequations}
  \begin{align}
    [L_n, L_m] &= (n-m) L_{n+m}, \quad [\bar{L}_n, \bar{L}_m] = (n-m) \bar{L}_{n+m}, \\
    [L_n, M_{r,s}] &= \left( \frac{n+1}{2} - r \right) M_{n+r, s}\,, \quad [\bar{L}_n,M_{r,s}] = \left( \frac{n+1}{2} - s \right) M_{r,n+s}.
\end{align}  
\end{subequations}
The $M_{r,s}$ are the supertranslations in the above and the $L_n, \bar{L}_n$ represent the superrotations. The isomorphism between $\mathfrak{CCarr}_3$ and $\mathfrak{ebms}_4$ is at the heart of the recent resurgence of the construction of 4D flatspace holography using Carrollian symmetries \cite{Bagchi:2022emh, Donnay:2022aba}. 

\medskip

Given the fact that our Lagrangian \eqref{L2} is already invariant under infinite supertranslatations, we check for the transformation of the action under infinite superrotations, where now we take the null time direction to be $x^-$. The spinor transforms as
\begin{subequations}
    \begin{align}
    \delta_{L_n}\Psi &= \left[z^{n+1}\partial_z + \frac{1}{2}(n+1)z^n\left(x^-\partial_- + \Delta \right)\right]\Psi - (n+1)z^n J\Psi \\
    \delta_{L_n}\bar\Psi &= \left[z^{n+1}\partial_z + \frac{1}{2}(n+1)z^n\left(x^-\partial_- + \Delta \right)\right]\bar\Psi + (n+1)z^n\bar{\Psi}J
\end{align}
\end{subequations}
where $J$ is defined in \eqref{spingen}. The Lagrangian transforms as

\begin{align}
    \delta_{L_n}\mathcal{L} = i\partial_z(z^{n+1}\mathcal{L}) + \frac{i}{2}(n+1)z^n\partial_-(x^-\mathcal{L})  + \frac{i}{2}(n+1)z^n(2\Delta -2)\mathcal{L}
\end{align}
 The above action is invariant for $\Delta=1$, which matches with our earlier analysis (see around Eq.\eqref{scaling}). One can similarly check the invariance under $\bar{L}_n$. We thus see that our action enjoys invariance under the full extended BMS$_4$ algebra.

\subsubsection{Discrete symmetries}
Discrete symmetries form an important part of all fermionic field theories. Here we investigate how parity, time-reversal and charge conjugation act on these lightcone fermions. We will find some interesting features here. 

\begin{itemize}
    \item {\textbf{Parity.}} The two-component spinors $\psi_{\pm}$ are related to each other by parity transformation. The Parity transformation in Minkowski spacetime defined by,
\begin{eqnarray}
    t \to t\,, \quad x^a \to -x^a
\end{eqnarray}
which in the lightcone coordinate, acts as
\begin{eqnarray}
    x^{\pm} \to x^{\mp}\,,\quad x^i \to -x^i.
\end{eqnarray}
Under Parity, spatial rotation remains invariant, while boosts change sign. In the relativistic theory, parity exchanges the two Weyl components of a Dirac spinor, 
\begin{eqnarray}
    \mathbf{P}: \psi_{\pm}(t,\vec{x}) \to \psi_{\mp}(t,-\vec{x})
\end{eqnarray}
In the Carrollian setting, parity acts more subtly. Since the two Carroll subalgebras are interchanged under $ x^{\pm} \to x^{\mp}$, parity maps the two spinor representations into each other. Explicitly, one finds
\begin{eqnarray}\label{carrparity}
    \mathbf{P}: \psi_{\pm}^{(1),(2)} \to \psi_{\mp}^{(2),(1)}.
\end{eqnarray}
As a consequence, the equations of motion are mapped into each other under parity
\begin{subequations}
 \begin{eqnarray}
    &&i\sigma\partial_-\psi_+^{(1)} = 0 \xrightarrow{\mathbf{P}} i\sigma\partial_+\psi_-^{(2)} = 0\\
    &&i\tilde{\sigma}\partial_-\psi_-^{(1)} = 0 \xrightarrow{\mathbf{P}} i\tilde{\sigma}\partial_+\psi_+^{(2)} = 0
\end{eqnarray}   
\end{subequations}
Parity therefore relates the two Carroll fermion theories obtained from opposite null contractions, setting their physical equivalence. 

\item {\textbf{Time-reversal.}} Under time reversal $T: t\to -t\,, x^i \to x^i$ (and hence $x^{\pm} \to -x^{\mp}$), the same exchange of null directions occurs, thus relating the two Carrollian theories in the same manner as parity. 

\item {\textbf{Charge conjugation.}} We define the charge conjugate of spinor $\Psi$ as 
\begin{align}\label{cc}
    \Psi^{(c)} =  C\Psi^* 
\end{align}
Here $C$ is a $4\times 4$ matrix, satisfying 
\begin{align}
    C^\dagger C = 1\,,\quad (\g_{a})^* = -C^\dagger\g_a C
\end{align}
It can be checked that  
\begin{equation}
    C =\begin{pmatrix}
        0 & \sigma^2\\
        \sigma^2 & 0
    \end{pmatrix}
\end{equation}
Also, $(S_{ab})^* = C^\dagger S_{ab}C$, which validates the definition of charge conjugation \eqref{cc} under Carroll transformation (spin part). Charge conjugation acts in the standard way and remains a symmetry of the Carrollian action.

We now define the Carroll equivalent of the \emph{Majorana condition} on the spinors. This is defined by requiring the spinor to be equal to its charge conjugate,
\begin{equation}
\Psi^{(c)} = \Psi \quad \Longleftrightarrow \quad C \Psi^* = \Psi \,.
\end{equation}
Thus in the two-component notation, the Majorana condition implies
\begin{equation}\label{Majcond}
\psi_+ = \sigma^2 \psi_-^*, 
\qquad
\psi_- = \sigma^2 \psi_+^* \,,
\end{equation}
showing that the two Weyl components are not independent.

The compatibility of this condition with the Carrollian equations of motion \eqref{carrLeom} needs to be verified. Substituting \eqref{Majcond} into the first equation gives
\begin{equation}
i\sigma \partial_- (\sigma^2 \psi_-^*) = 0 \,.
\end{equation}
Using the identity
\begin{equation}
\sigma \sigma^2 = \sigma^2 \tilde{\sigma}^{\,T},
\end{equation}
and taking complex conjugation and after left multiplication by $\sigma^2$, this reduces to
\begin{equation}
i\tilde{\sigma} \partial_- \psi_- = 0 \,,
\end{equation}
which is precisely the second equation of motion. Hence, the Majorana condition is consistent with the Carrollian dynamics.
\end{itemize}
Having considered continuous and discrete spacetime symmetries, we will now move to other symmetries of our contracted Dirac action.

\subsubsection{Other symmetries}

\begin{itemize}
    \item \textbf{Internal symmetries.} The action \eqref{CarrL} is invariant under a global $U(1)$ phase rotation,
\begin{equation}
\Psi \to e^{i\theta}\Psi , \qquad
\bar{\Psi} \to \bar{\Psi} e^{-i\theta} ,
\end{equation}
corresponding to fermion number conservation. Moreover, since the two chiral sectors $\psi_+$ and $\psi_-$ decouple in the Carrollian limit (in this particular basis), the symmetry is enhanced to
\begin{equation}
\psi_+ \to e^{i\theta_+}\psi_+ , \qquad
\psi_- \to e^{i\theta_-}\psi_- ,
\end{equation}
yielding an internal $U(1)_+ \times U(1)_-$ symmetry.

\item \textbf{Null dilatation.} Perhaps the most intriguing and insightful of the symmetries is the action of the generator $B$ \eqref{poi-lc}, the Lorentz boost along the $z$-direction. In gamma-matrix language, we referred to this as $S_{tz}$. We saw that this acts similar to a dilatation generator that rescales the two null directions oppositely \eqref{B-act}, which in terms of coordinates gives
\begin{equation}
    x^+ \to \lambda x^+, \qquad x^- \to \lambda^{-1} x^- .
\end{equation}
On spinors, the corresponding generator is given by
\begin{equation}
V \equiv S_{tz} = S_{+-}
= \frac{1}{4}[\gamma_+,\gamma_-]= \frac{1}{2}\begin{pmatrix}
        \sigma^3 & 0\\
         0 & -\sigma^3
    \end{pmatrix} ,
\end{equation}
which acts diagonally on the chiral components. The generator $V$ satisfies the following commutation relations
\begin{eqnarray}
    [V,S_{-i}] = +S_{-i} \,,\qquad [V,S_{+i}] = -S_{+i}\,,\qquad [V,J] = 0
\end{eqnarray}
One can introduce the projection operator
\begin{eqnarray}
    P_{\pm} = \frac{1}{2} \pm V
\end{eqnarray}
which satisfy the standard properties $P_{\pm}^2 = P_{\pm}$ and $P_+P_- = 0$. The generator $V$ seems like the analogue of the relativistic $\gamma^5$ which defines the chirality operators $\hat{P} = \frac{1}{2}(1\pm \gamma^5)$ in the relativistic fermionic theory. 

The lightcone projection operators act on the spinor as 
\begin{subequations}\label{projspin}
    \begin{eqnarray}
    &&P_+\Psi  = P_+\begin{pmatrix}
        \psi_+\\
        \psi_-
    \end{pmatrix} = \begin{pmatrix}
        \sigma\psi_+\\
        \tilde{\sigma}\psi_-
    \end{pmatrix} = \begin{pmatrix}
        \psi_{+A}\\
        0\\
        0\\
        \psi_{-B}
    \end{pmatrix},\\
    &&P_-\Psi  = P_-\begin{pmatrix}
        \psi_+\\
        \psi_-
    \end{pmatrix} = \begin{pmatrix}
        \tilde{\sigma}\psi_+\\
        \sigma\psi_-
    \end{pmatrix} = \begin{pmatrix}
        0\\
        \psi_{+B}\\
        \psi_{-A}\\
        0
    \end{pmatrix}.
\end{eqnarray}
\end{subequations}
This decomposition makes manifest that the Carrollian fermions naturally split into sectors of definite null scaling, governed by the generator $V$. This is also consistent with the discussion of \eqref{rep1} and \eqref{rep2}, where it was shown that $(1,4)$ components of spinor $\Psi$ live on one null surface while the $(2,3)$ lies on the other.

\end{itemize}

\subsection{Propagators}
As a first foray into quantum aspects of fermions in the lightcone and their Carrollian structure, we analyse the structure of the propagator. We will compute the propagator directly from the Lagrangian \eqref{CarrL}. Since the two sectors are completely decoupled, we treat them separately. We begin with $\psi_+$ sector. Writing the spinor in components as  $\psi_+ = (\psi_{+A},\psi_{+B})^T$ and 
$\sigma = \mathrm{diag}(1,0)$, the kinetic term becomes
\begin{equation}
    \psi_+^\dagger\,\sigma\,\partial_-\psi_+ 
    = \begin{pmatrix}
    \psi_{+A}^* & \psi_{+B}^*
    \end{pmatrix}
      \begin{pmatrix}
      \partial_-\psi_{+A}\\0
      \end{pmatrix}
    = \psi_{+A}^*\,\partial_-\psi_{+A}\,.
\end{equation}
The component $\psi_{+B}$ drops out of the action. The action reduces to
\begin{equation}
    S_1 = -i\sqrt{2}\int dx^- d^2x^i\,\psi_{+A}^*\,\partial_-\psi_{+A}\,.
\end{equation}
We now Fourier transform the dynamical field,
\begin{equation}
    \psi_{+A}(x^-,x^i) 
    = \int\frac{dk^+d^2k^i}{(2\pi)^3}\,
    e^{-ik^+x^-+ik^ix_i}\,\tilde{\psi}_{+A}(k^+,k^i)\,.
\end{equation}
Substituting into the action $S_-$ and performing the position space integral, we find
\begin{equation}
    S_1 = \int\frac{dk^+d^2k^i}{(2\pi)^3}\,
    \tilde{\psi}_{+A}^*\left(-\sqrt{2}\,k^+\right)\tilde{\psi}_{+A}\,.
\end{equation}
The kinetic operator is linear in $k^+$ and inverting this we get the propagator
\begin{equation}
    G_{+A}(k^+) = \frac{i}{-\sqrt{2}\,k^++i\varepsilon}\,.
\end{equation}
A completely analogous analysis applies to $\psi_-$ sector. In this case $\psi_{-B}$ is dynamical and $\psi_{-A}$ drops out of the action. The corresponding propagator is, 
\begin{equation}
    G_{-B}(k^+) = \frac{i}{-\sqrt{2}\,k^++i\varepsilon}\,.
\end{equation}
Combining both sectors, the full momentum space propagator can be written in matrix form as
\begin{align}
    S_F^{\text{Carroll}}(k^+) = \frac{i}{-\sqrt{2}\,k^++i\varepsilon}
    \begin{pmatrix}
    \sigma & 0 \\ 
    0 & \tilde\sigma
    \end{pmatrix},
\end{align}
where the projectors $\sigma$ and $\tilde\sigma$ enforce propagation only in the dynamical subspaces of the $\psi_+$ and $\psi_-$ sectors respectively. 

\noindent
One can find its position space realization by Fourier transforming the above equation
\begin{align}
    S_F^{\text{Carroll}}(x,x')  = \int\frac{dk^+d^2k^i}{(2\pi)^3}\,e^{-ik^+(x^--x'^{-})+ik^i(x_i-x_i')}\,
    S_F^{\text{Carroll}}(k^+)\,.
\end{align}
Since the propagator is independent of $k^i$, the integral yields
\begin{align}
    \int\frac{d^2k^i}{(2\pi)^2}\,e^{ik^i(x_i-x_i')} 
    = \delta^{(2)}(x^i-x'^{i})\,.
\end{align}
The remaining integral gives
\begin{align}
     \int\frac{dk^+}{2\pi}\,
    \frac{i\,e^{-ik^+(x^--x'^{-})}}{-\sqrt{2}\,k^++i\varepsilon} 
    = \frac{1}{\sqrt{2}}\,\theta(x^--x'^{-})\,.
\end{align}
The full position space propagator is 
\begin{align}
    S_F^{\text{Carroll}}(x,x') = \frac{1}{\sqrt{2}}\,\theta(x^--x'^{-})\,\delta^{(2)}(x^i-x'^{i})\begin{pmatrix}
    \sigma & 0 \\ 
    0 & \tilde\sigma
    \end{pmatrix}\,.
\end{align}
The delta function $\delta^{(2)}(x^i-x'^{i})$ reflects the ultra-locality of the theory in transverse directions and is a classical signature of Carrollian field theories. The step function $\theta(x^--x'^{-})$ signals unidirectional propagation along null direction $x^-$, and the chiral nature of the system.

For the theory that lives on $x^+$, a similar calculation yields identical results with $x^-$ replaced by $x^+$. The next steps in understanding fermions in this way would be the introduction of interaction terms and constructing the Feynman rules of the interacting theory and then moving on to computing amplitudes in the interacting theory.

\subsection{Connecting lightcone and Carroll expansions}
We now assemble the results of this section into an explicit connection between the lightcone spinor decomposition and the Carrollian fermionic theories of section \ref{sec2} and \ref{sec3}. 
\vspace{10 pt} 
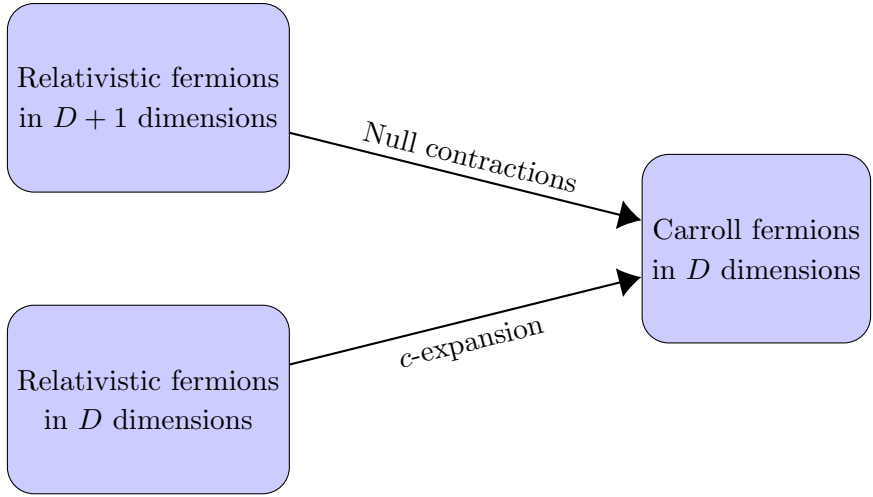
\begin{figure}[h!]
    \centering
    \begin{tikzpicture}[
    node distance=4cm,
    box/.style={
        draw,
        fill=blue!20,
        rounded corners=10pt,
        minimum width=2.5cm,
        minimum height=2.5cm,
        align=center
    },
    arrow/.style={
        -{Latex[length=3mm, width=4mm]},
        thick
    }
]

\node[box] (B) at (0,2) {Relativistic fermions\\ in $D+1$ dimensions};
\node[box] (A) at (0,-2) {Relativistic fermions\\ in $D$ dimensions};
\node[box] (C) at (8,0) {Carroll fermions\\ in $D$ dimensions};

\draw[arrow] (B) -- node[above, sloped] {Null contractions} (C);
\draw[arrow] (A) -- node[below, sloped] {$c$-expansion} (C);
\end{tikzpicture}

    \caption{The multi-dimensional flavour of Carroll fermions}
    \label{fig1}
\end{figure}

The lightcone gamma matrices $\g_{\pm}$ satisfy the degenerate Clifford algebra, which is precisely the Carroll Clifford algebra \eqref{CClf} with $h_{\mu\nu} = \mathrm{diag}(0,1,1)$. Under the null contractions $x^- \to \epsilon x^-$, the gamma matrix $\g^{-}$ (or $\g_+$), associated with the surviving null direction plays the role of the lower Carroll gamma matrix $\gc_0$. The light-cone fermion theory \eqref{CarrL} therefore realises the inhomogeneous representation of the Carroll Clifford algebra in the lower gamma sector. 

The two overlapping Carroll subalgebras identified in \eqref{carrsuba}, correspond to the two possible null contractions and thus two identical looking actions. Parity, as shown in \eqref{carrparity}, exchanges the two theories, confirming they are physically equivalent. The null dilatation $V$ and the projector $P_{\pm}$ decompose the spinor into two sectors \eqref{projspin}. The sector $P_+\Psi$ is acted upon non-trivially by the Carroll boosts of representation (1) in \eqref{rep1}, while $P_-\Psi$ is inert under those boosts. This flips upon considering Carroll boosts of representation (2) in \eqref{rep2}. This mirrors precisely the split between dynamical $(\phi)$ and non-dynamical $(\chi)$ components in the intrinsic lower gamma Carroll theory \eqref{lga}. In addition, the doubling of representation dimension in odd Carrollian spacetime directions admits a natural lightcone interpretation. 

Coming to the expansion method, the lower gamma Carroll theory corresponds to the leading-order $c$-expansion and to the null-contracted lightcone theory with the nilpotent $\gamma^-$; the upper gamma Carroll theory corresponds to the next-to-leading-order $c$-expansion. Both are realised in the imhomogeneous representation $\mathcal{R}_I$.

\section{Conclusions}\label{sec5}

\subsection{Summary of results}

In this paper, we have undertaken a systematic study of fermions in Carrollian geometry and tried to relate them to relativistic fermions through two different routes via Carroll-expansions in the same number of dimensions and in the lightcone formalism connecting $(D+1)$-dimensional relativistic fermions to $D$-dimensional Carroll fermions. 

We began by providing a lightning review of Carroll symmetry and fermions derived in an intrinsic Carrollian framework. We then discussed some subtle issues of representation of Carroll Clifford algebra specifically in odd spacetime dimensions. 

Complementing the intrinsic analysis, we derived Carrollian fermion theories from a relativistic Dirac action via a systematic expansion in the speed of light. By carefully analysing the scaling of spinor components, we demonstrated how both classes of intrinsic Carrollian actions emerge naturally from the relativistic parent theory. In particular, the homogeneous and inhomogeneous Carroll fermions can be traced back to different consistent scalings of relativistic spinor components, providing a unifying perspective on their origin.

In the next part of our paper, we analysed fermions formulated in light-cone coordinates and clarified their relation to Carrollian fermions. By rewriting the relativistic Dirac theory in a null basis, we showed that the light-cone decomposition naturally splits spinor components. We demonstrated that the Carroll limit can be understood as a null contraction of the light-cone theory, in which one of the light-cone directions becomes Carrollian time. In this formulation, the degeneracy of the Carroll metric is directly inherited from the null structure of Minkowski spacetime, and the resulting Clifford algebra is realised by nilpotent gamma matrix. The light-cone gamma matrices therefore provide a concrete and physically motivated representation of the Carroll Clifford algebra, rather than an abstract algebraic construction. An important outcome of this analysis is the precise relation between light-cone and intrinsic Carrollian representation. The light-cone component naturally maps onto the inhomogeneous Carroll representation. 

Taken together, these results show that Carrollian fermions arise in a natural way from relativistic fermions, either through $c$-expansions or via null contractions. This viewpoint provides a useful bridge between intrinsic Carrollian field theories and light-cone dynamics and may prove valuable in applications in the context of condensed matter physics, specifically in systems with flat bands.

\subsection{Discussions and future directions}
Our current explorations into Carroll fermions paves the way for many further developments. We list some of these below. 

\begin{itemize}

\item{\em Massive theories: } Throughout this work we have restricted attention to massless fermions. It is natural to ask whether how a Carroll invariant mass term can be systematically obtained from $c$-expansion. While we did not address this in the main text, we provide some preliminary remarks in Appendix \ref{appE}. 

\item {\em Quantum aspects:} We took our first steps towards a quantum theory of fermions while constructing the propagators in the previous section and this is a project we wish to pursue in detail. It has been speculated in the literature that quantizing Carrollian theories is fraught with danger \cite{deBoer:2023fnj, Cotler:2024xhb, Cotler:2025dau}. We, however, do not share this sentiment. An important point here, as we have now shown even for fermionic theories, is that Carrollian theories can be understood in terms of relativistic theories in the lightcone. So it is natural to use age-old techniques of lightcone field theories to understand quantum Carrollian theories. Also, some of the problems in dealing with Carroll theories arise from sending the speed of light strictly to zero. The $c$-expansion, on the other hand, seems to be a better tool in this regard and should hold the key to regulating Carrollian theories. We plan to return to these aspects in the near future, specifically to the quantization of the fermionic theories. 

\item {\em Gauge theories with fermionic matter}: A natural question is to explore Carrollian gauge theories coupled to fermionic matter. These would include Carroll Quantum Electrodynamics, Carrollian Yang Mills and Chern-Simons theories with fermions. For some previous explorations in these directions, see \cite{Bagchi:2019xfx, Bagchi:2019clu, Basu:2018dub, Islam:2023rnc, deBoer:2021jej}. The new insights about fermions should be very useful in understanding these Carrollian theories. 

\item {\em Carrollian Supersymmetry}: Various attempts at Carrollian supersymmetry have been made in the recent past. See e.g. \cite{Bagchi:2022owq, Koutrolikos:2023evq, Kasikci:2023zdn, Zorba:2024jxb, Concha:2024dap, Zheng:2025cuw, Bruce:2026yvw, Bulunur:2026yav, Ergec:2026baz}. However the intricate structure of Carrollian fermions has not been fully appreciated before. This should now help in building Carrollian SUSY from scratch without having to resort primarily to limits.  

\item {\em Flat holography and fermions -- ABJM}: One of the principle goals of the construction of Carrollian theories is to address questions of holography in asymptotically flatspace through the Carrollian holography programme. An important route to flat holography has been attempting a flat-space limit of known holograms in AdS. Prominent among them is the AdS$_4$/CFT$_3$ correspondence which connects M-theory on AdS$_4 \times$S$^7/\mathbb{Z}_k$ to 3D $\mathcal{N}=6$ Superconformal Chern-Simons theory or ABJM theory on the boundary \cite{Aharony:2008ug}. The Carrollian limit of ABJM would crucially contain fermions and in upcoming work \cite{ABJM}, we address this carefully to build a Carrollian ABJM with matter fields building on earlier work \cite{Bagchi:2024efs, Lipstein:2025jfj}. This work lays the groundwork for building a concrete flatspace hologram with Carrollian superconformal symmetries. 

\item {\em{Carrollian $\mathcal{N} =4$ Super Yang-Mills:}} The most celebrate of the holographic correspondences is of course Maldacena's original AdS$_5$/CFT$_4$ correspondence \cite{Maldacena:1997re} which relates Type IIB superstring theory on AdS$_5 \times$S$^5$ to $\mathcal{N}=4$ Super Yang-Mills in $D=4$. Building a proper Carrollian version of $\mathcal{N}=4$ SYM is one of our important future goals. 
\end{itemize}

Clearly, this paper builds the basics of Carrollian fermions which we believe would be very important for a wide range of applications in various branches of theoretical physics where Carrollian symmetries are finding central roles. We hope to return to the various questions listed above and others in the near future. 

\section*{Acknowledgement}
We thank Aritra Banerjee, Rudranil Basu, Joydeep Chakrabortty, Sharang Iyer, Arthur Lipstein,  Nachiketh M, Pushkar Soni, Alex Jiayi Zhang for discussions. 
AB was supported partially by a Swarnajayanti Fellowship from Anusandhan National Research Foundation (ANRF) under grant SB/SJF/2019-20/08 and also by an ANRF grant
CRG/2022/006165. SM is supported by Fellowship for Academic and Research Excellence (FARE) at IIT
Kanpur.

\bigskip
\bigskip

\newpage

\appendix

\section*{APPENDICES}
\section{Basics: Spinors of 4D Lorentz algebra}\label{appA}
To set up conventions and nomenclature, we will briefly some basics of Lorentzian spinors in $D=4$ spacetime as a warm up for our discussions of spinors on the lightcone. The relativistic Clifford algebra is defined by
\begin{equation}
    \{\g^{\mu},\g^{\nu}\} = 2\eta^{\mu\nu}I
\end{equation}
with $\mu = 0,1,2,3$. Our metric convention is $\eta^{\mu\nu} = \text{diag}(-1,1,1,1)$. There exist many representations which satisfy the above Clifford algebra. We adopt the \textit{chiral} or \textit{Weyl} representation 
\begin{eqnarray}\label{Weylrel}
    \gamma^0 = \begin{pmatrix}
        0 & \sigma^0\\
        -\sigma^0 & 0
    \end{pmatrix}\qquad
    \gamma^i = \begin{pmatrix}
        0 & \sigma^i\\
        \sigma^i & 0
    \end{pmatrix},
\end{eqnarray}
where, $\sigma^0$ is the $2\times 2$ identity matrix and $\sigma^i$'s are the usual Pauli spin matrices obeying
\begin{eqnarray}
    \{\sigma^i,\sigma^j\} = 2\delta^{ij}\,,\qquad [\sigma^i,\sigma^j] = 2i\epsilon^{ijk}\sigma^k\,.
\end{eqnarray}
From these $\g$ matrices, one can construct a new set of matrices defined as
\begin{eqnarray}
    S^{\mu\nu} = \frac{1}{4}[\g^{\mu},\g^{\nu}]
\end{eqnarray}
which induce a representation of Lorentz algebra
\begin{eqnarray}
    [S^{\mu\nu},S^{\rho\sigma}] = \eta^{\nu\rho}S^{\mu\sigma} - \eta^{\mu\rho}S^{\nu\sigma} + \eta^{\mu\sigma}S^{\nu\rho} -
    \eta^{\nu\sigma}S^{\mu\rho}\,.
\end{eqnarray}
Here we have three Lorentz boosts $S^{0i}$ and three spatial rotations $S^{ij}$, where $\{i,j\} = 1,2,3$. In the \textit{Weyl} representation, they take block-diagonal form:
\begin{eqnarray}\label{lor4d}
    S^{0i} = \frac{1}{2}\begin{pmatrix}
        \sigma^i & 0\\
         0 & -\sigma^i
    \end{pmatrix}\qquad
    S^{ij} = \frac{i}{2}\epsilon^{ijk}\begin{pmatrix}
        \sigma^k & 0\\
        0 & \sigma^k
    \end{pmatrix}.
\end{eqnarray}
Consider a Dirac spinor $\Psi^{\alpha}(x)$ with four complex components $(\alpha = 1,2,3,4)$. Under Lorentz transformation, it transforms as
\begin{eqnarray}
    \Psi^{\alpha}(x) \to S[\Sigma]^{\alpha}_{\,\beta}\,\Psi^{\beta}(\Sigma^{-1}x)
\end{eqnarray}
with $S[\Sigma] = \exp{\left(\frac{1}{2}\Omega_{\mu\nu}S^{\mu\nu}\right)}$ and $\Sigma = \exp{\left(\frac{1}{2}\Omega_{\mu\nu}J^{\mu\nu}\right)}$, where $J^{\mu\nu}$ are also Lorentz generators in coordinate basis.
For rotations, with $\Omega_{ij} = \epsilon_{ijk}\varphi^k$,
\begin{eqnarray}
    S[\Sigma_{rot}]= \exp{\left(\frac{1}{2}\Omega_{ij}S^{ij}\right)} = \begin{pmatrix}
        e^{i\vec{\varphi} \cdot \vec{\sigma}/2} & 0\\
        0 & e^{i\vec{\varphi} \cdot \vec{\sigma}/2}
    \end{pmatrix}.
\end{eqnarray}
For Lorentz boosts, with $\Omega_{i0} = -\Omega_{0i} = \vartheta_i$,
\begin{eqnarray}
    S[\Sigma_{boost}] = \exp{\left(\frac{1}{2}\Omega_{0i}S^{0i}\right)} = \begin{pmatrix}
        e^{-\vec{\vartheta} \cdot \vec{\sigma}/2} & 0\\
        0 & e^{\vec{\vartheta} \cdot \vec{\sigma}/2}
    \end{pmatrix}.
\end{eqnarray}
Both transformations are block-diagonal, so this representation is reducible. Hence, one can decompose the four-component spinor into two-component \textit{Weyl} spinors, 
\begin{eqnarray}\label{2cw}
    \Psi= \begin{pmatrix}
        \psi_+ \\
        \psi_-
    \end{pmatrix}.
\end{eqnarray}
Under rotations, they transform as
\begin{eqnarray}
    \psi_{\pm} \to e^{i\vec{\varphi} \cdot \vec{\sigma}/2}\,\psi_{\pm}, 
\end{eqnarray}
whereas, under Lorentz boosts, they transform as
\begin{eqnarray}
    \psi_{\pm} \to e^{\mp \vec{\vartheta} \cdot \vec{\sigma}/2}\,\psi_{\pm}\,.
\end{eqnarray}
Hence $\psi_+$ sits in the $(\frac{1}{2},0)$ representation of the Lorentz group, while $\psi_-$ sits in the $(0,\frac{1}{2})$ representation. 

\medskip

\noindent Now, we come to the Dirac action, which is given by,
\begin{eqnarray}
    \mathcal{S} = \int d^4x\,\bar\Psi(x)\left(i\g^{\mu}\p_{\mu} - m\right) \,\Psi(x)\,.
\end{eqnarray}
The adjoint $\bar\Psi$ is defined as $\bar\Psi = \Psi^\dagger\g^0$. We have also included the Dirac mass term $m\bar{\Psi}\Psi$. 
Under the decomposition \eqref{2cw}, the action becomes
\begin{eqnarray}\label{Weylact}
    \mathcal{S} = \int d^4x\, \left(i\psi_+^\dagger \sigma^{\mu}\p_{\mu}\psi_+ + i\psi_-^\dagger \bar{\sigma}^{\mu}\p_{\mu}\psi_- - im\left(\psi_+^\dagger\psi_- - \psi_-^\dagger\psi_+\right) \right)\,.
\end{eqnarray}
where
\begin{eqnarray}
    \sigma^{\mu} = (\sigma^0, \sigma^i)\,,\quad \bar{\sigma}^{\mu} = (\sigma^0, -\sigma^i)\,.
\end{eqnarray}
Symmetries of \eqref{Weylact} are quite different for the massless and massive case. For massless one, the action exhibits $U(1) \times U(1)$ symmetry under which the Weyl spinors transform as $\psi_+ \to e^{ia}\psi_+$ and $\psi_- \to e^{ib}\psi_-$, while for the massive case this is the symmetry only when $a=b$. 
In the presence of mass term, one cannot decouple the spinors. However, for massless case, they can be decoupled and we are left with the equation of motion 
    \begin{eqnarray}
    i\sigma^{\mu}\p_{\mu}\psi_+ = 0,\quad 
    i\bar{\sigma}^{\mu}\p_{\mu}\psi_- = 0
\end{eqnarray}
These are the \textit{Weyl} equations. 

\medskip

\noindent The above discussion relied on the fact that in the Weyl representation, the Lorentz generators are block-diagonal. However, in general representations, chirality is defined via 
\begin{eqnarray}
    \g^5 = -i\g^0\g^1\g^2\g^3
\end{eqnarray}
which satisfies 
\begin{eqnarray}
    (\g^5)^\dagger = \g^5\,,\qquad (\g^5)^2 = I\,,\qquad \{\g^5,\g^\mu\} = 0\,.
\end{eqnarray}
Since $[\g^5,S^{\mu\nu}] = 0$, the representation is reducible in any basis. Using $\g^5$, one can construct projection operators $P_{\pm} = \frac{1}{2}(1\pm \g^5)$,  such that $P_{\pm}^2 = P_{\pm}$, $P_+P_- = P_-P_+ = 0$. These define Weyl spinors intrinsically via $\psi_{\pm} = P_{\pm}\Psi$. In chiral basis,
\begin{eqnarray}
    \g^5 =\begin{pmatrix}
         1 & 0\\
         0 & -1
    \end{pmatrix}\,.
\end{eqnarray}
so the projectors simply select the upper and lower two components. 

\medskip

A complex Dirac spinor has 8 real degrees of freedom. Each Weyl spinor has 4 real d.o.f. The first-order Weyl equations impose 2 real constraints per spinor yielding 2 physical d.o.f. per Weyl spinor -- corresponding to a massless helicity-$1/2$ particle with $2$ states. The massive Dirac field has 4 physical d.o.f. matching a spin-$1/2$ particle. 

\section{Dimension of Carroll Spinor in odd dimensions}\label{appB}
For non-degenerate metric, the Clifford algebra is semi-simple and its irreducible representation has fixed dimension $2^{[D/2]}$. However, when the metric is degenerate, the algebra is no longer semi-simple and admits inequivalent representations of different dimensions. While a two-component realization can formally satisfy the degenerate Clifford algebra, it necessarily renders the nilpotent generator trivial and therefore fails to provide a good theory. \\
{\bf Claim:} In three-dimensional Carrollian spacetime, any faithful representation of the Carroll Clifford algebra with a non-vanishing nilpotent gamma matrix $\tilde{\gamma}_0$ has minimal dimension four. More generally, in odd spacetime dimensions $D=2N+1$, the minimal dimension of a Carroll Clifford representation is $2^{N+1}$.\\
{\bf Proof:} We first consider the case of three spacetime dimensions. The Carroll Clifford algebra is generated by matrices $\{\tilde{\gamma}_0,\tilde{\gamma}_1,\tilde{\gamma}_2\}$,
satisfying the relations
\begin{equation}\label{cr3}
\{\tilde{\gamma}_0,\tilde{\gamma}_0\}=0, \qquad
\{\tilde{\gamma}_0,\tilde{\gamma}_i\}=0, \qquad
\{\tilde{\gamma}_i,\tilde{\gamma}_j\}=2\delta_{ij}, \quad i,j=1,2.
\end{equation}
In particular, $(\tilde{\gamma}_0)^2=0$, so $\tilde{\gamma}_0$ is nilpotent. The spatial gamma matrices $\tilde{\gamma}_1,\tilde{\gamma}_2$ generate the Euclidean Clifford algebra $\mathrm{Cl}(2)$, whose irreducible complex representation is two-dimensional. Hence, if a two-dimensional representation of the full Carroll Clifford algebra were to exist, the matrices $\tilde{\gamma}_1$ and $\tilde{\gamma}_2$ would necessarily act irreducibly on the spinor space $V\simeq \mathbb{C}^2$. However, in a two-dimensional irreducible representation of $\mathrm{Cl}(2)$, the set of matrices generated by $\tilde{\gamma}_1$ and $\tilde{\gamma}_2$ already spans the full matrix algebra. There exists no {\bf non-zero} nilpotent matrix that anticommutes with both generators of $\mathrm{Cl}(2)$ while preserving their Clifford relations. Consequently, a two-dimensional representation cannot realise the full Carroll Clifford algebra in three dimensions.

The minimal resolution of this obstruction is to double the representation space,
\begin{equation}
V = V_1 \oplus V_2 , \qquad \dim V_{1,2} = 2,
\end{equation}
and choose a block representation
\begin{equation}
\tilde{\gamma}_i =
\begin{pmatrix}
\sigma^i & 0 \\
0 & -\sigma^i
\end{pmatrix}, \qquad
\tilde{\gamma}_0 =
\begin{pmatrix}
0 & 0 \\
\mathbb{I}_2 & 0
\end{pmatrix},
\end{equation}
where $\sigma^i$ are Pauli matrices. One verifies directly that these matrices satisfy all Carroll Clifford relations. Therefore, the minimal faithful representation in three spacetime dimensions is four-dimensional. 

\medskip

However, if we stick to homogeneous representation, then one can have following choice of representation
\begin{align}
    \tilde{\gamma}_0 = 0\,,\quad \tilde{\gamma}_i = \sigma_i\,,
\end{align}
which satisfies the Clifford algebra in 3D \eqref{cr3}.
In this case, gamma matrices admit a minimal realisation in terms of $2 \times 2$ matrices. 

\section{c-expansion of field theory}\label{appC}
\subsection*{Scalar field}
We start with a relativistic real scalar field $\varphi$ in D-dimensional Minkowksi spacetime , with the metric signature $(-,+,+,...)$. The action is given by 
\begin{align}\label{KGaction}
    \mathcal{S}_{rel} = \int d^Dx\,\mathcal{L}_{rel} &= \int\, d^Dx\,\left(\frac{1}{2}\eta^{\mu\nu}\partial_{\mu}\varphi\partial_{\nu}\varphi - V(\varphi)\right)\nonumber\\ &= \int\, dtd^{D-1}x\,\left(-\frac{1}{2c^2}(\partial_t\varphi)^2 + \frac{1}{2}(\partial_i\varphi)^2 + V(\varphi)\right), 
\end{align}
where $V(\varphi)$ is a potential term. We have also explicitly reinstated the speed of light $c$ in the time derivative term to track its behaviour in the Carrollian expansion \cite{deBoer:2021jej}. To perform a systematic expansion in small $c$, we assume that up to an overall power of $c$, the field $\varphi$ admits a regular analytic expansion around $c=0$:
\begin{align}
    \varphi = c^\alpha\left(\varphi_0 + c^2\varphi_1 + c^4\phi_2 + ... \right)
\end{align}
for some exponent $\alpha$. This naturally implies the Lagrangian itself can be organized in a similar fashion,
\begin{align}
    \mathcal{L} = c^{\hat{\alpha}}\left( \mathcal{L}_0 +  c^2\mathcal{L}_1 +  c^4\mathcal{L}_2 + ... \right).
\end{align}
Substituting the field expansion into \eqref{KGaction}, we obtain the LO and NLO contributions as 
\begin{align}
    \mathcal{L}_0 = \frac{1}{2}(\partial_t\varphi_0)^2\,,\qquad \mathcal{L}_1 = \partial_t\varphi_0\partial_t\varphi_1 - \frac{1}{2}(\partial_i\varphi_0)^2. 
\end{align}
Now we examine how Lorentz transformations get affected in this expansion. Spatial rotations contain no explicit  $c$-factor, and thus act identically at all orders. In contrast, Lorentz boosts introduce relative factor of $c$. For a scalar field $\varphi$, the infinitesimal boost transformation reads,
\begin{eqnarray}\label{relboost}
    \delta_{B}\varphi = ct\beta^i\partial_i\varphi + \frac{1}{c}\beta^ix_i\partial_t\varphi\,,
\end{eqnarray}
where $\beta^i$ is the Lorentz boost parameter. To obtain a smooth Carrollian limit, we rescale $\beta^i = c\,b^i$, with $b^i$ the Carroll boost parameter. One can then substitute the field expansion in \eqref{relboost}, and find the transformations for LO and NLO fields as
\begin{eqnarray}
    \delta_{B}\varphi_0 = b^ix_i\partial_t\varphi_0\,, \qquad \delta_{B}\varphi_1 = b^ix_i\partial_t\varphi_1 + tb^i\partial_i\varphi_0\,.
\end{eqnarray}

It is straightforward to check that the Lagrangian $\mathcal{L}_0$ is Carroll boost invariant, while the Lagrangian $\mathcal{L}_1$ is not. However, one can make it Carroll boost invariant by adding a Lagrange multiplier to $\mathcal{L}_1$, which sets $\partial_t\varphi_0$ to zero on-shell.

\subsection*{Gauge field}
Now we consider Maxwell's action $\mathcal{L} =- \frac{1}{4}F_{\mu\nu}F^{\mu\nu}$ and study its Carrollian limit via c-expansion \cite{deBoer:2021jej}. A key feature of this expansion is the unequal treatment of temporal and spatial components of the gauge field $A_{\mu}$. This asymmetry is intrinsic to the Carrollian contraction \cite{Duval:2014uoa} and reflects the unequal footing of space and time. This distinction is most transparent when the gauge field is viewed as a one-form
\begin{eqnarray}
    A = A_{\mu}dx^{\mu} = A_0dx^0 + A_idx^i = A_0cdt + A_idx^i = A_tdt + A_idx^i\,,
\end{eqnarray}
where we defined $A_t = cA_0$, And the expansion around $c=0$ is given by,  
\begin{eqnarray}
    A_t = c^\alpha\sum_{n = 0}^{\infty}\left(A_t^{(n)}c^{2n}\right),~~\qquad A_i = c^\alpha\sum_{n = 0}^{\infty}\left(A_i^{(n)}c^{2n}\right),
\end{eqnarray}
where $\alpha$ is some scaling exponent. Substituting these expressions into the Maxwell action, and organizing terms order by order in $c$, we obtain the leading and sub-leading Lagrangians as
\begin{eqnarray}
  &\mathcal{L}_0 = \frac{1}{2}\left(E_i^{(0)}\right)^2,\label{LOem} \\~~
  &\mathcal{L}_1 = E_i^{(0)}E_i^{(1)} - \frac{1}{4}\left(F_{ij}^{(0)}\right)^2.\label{NLOem}
\end{eqnarray}
Here, $E_i^{(n)} = cF_{0i}^{(n)} = \partial_tA_i^{(n)} - \partial_iA_t^{(n)}$ and $B_i^{(n)} = \frac{1}{2}\epsilon_{ijk}F_{jk}^{(n)}$ are electric and magnetic fields at each order. We note that the leading-order action $\mathcal{L}_0$ in \eqref{LOem} coincides with the result obtained by directly taking the Carrollian limit of Maxwell theory, confirming the consistency of the expansion. 

We now examine the realization of Carrollian boosts in the c-expansion. Under an infinitesimal boost with parameter $\beta^i$, the relativistic gauge field transforms as
\begin{equation}
    \delta_C A_\mu
    = c t \beta^i \partial_i A_\mu
    + \frac{1}{c} \beta^i x_i \partial_t A_\mu
    + \hat{\delta} A_\mu ,
\end{equation}
with intrinsic variations
\begin{equation}
    \hat{\delta} A_0 = \beta_i A_i,
    \qquad
    \hat{\delta} A_i = \beta_i A_0 .
\end{equation}
Substituting the Carrollian expansions of $A_t$ and $A_i$, we find that the unequal scaling of the temporal and spatial components leads to a nontrivial mixing between different orders in the $c$-expansion. The resulting transformations of the $n$-th order fields are
\begin{align}
    \delta A_t^{(n)} &=
    b^i x_i \partial_t A_t^{(n)}
    + t b^i \partial_i A_t^{(n-1)}
    + b^i A_i^{(n-1)}, \\
    \delta A_k^{(n)} &=
    b^i x_i \partial_t A_k^{(n)}
    + t b^i \partial_i A_k^{(n-1)}
    + b^i \delta_{ik} A_t^{(n)} ,
\end{align}
where $A_\mu^{(-1)} = 0$ by convention. 

Varying the leading-order Lagrangian \eqref{LOem} with respect to $A_t$ and $A_i$ yields the Carrollian analogues of Gauss’s law and Ampère’s law at leading and subleading orders. The remaining equations of motion follow from the Bianchi identity,
\begin{equation}
    \epsilon^{\mu\nu\rho\sigma} \partial_\nu F_{\rho\sigma} = 0 ,
\end{equation}
which decomposes, order by order in $c$, into
\begin{equation}
    \partial_i B_i^{(n)} = 0,
    \qquad
    \partial_t B_i^{(n)} + (\nabla \times E^{(n)})_i = 0 .
\end{equation}

\section{Verification of LO and NLO fermionic action}\label{appD}
Consider the action given by \eqref{LO}
\begin{align}
    \mathcal{L}_{(0)} = i\phi_0^\dagger\dot{\phi}_0
\end{align}
Under the transformation \eqref{lobt}, the Lagrangian transforms as
\begin{align}
    \delta_C\mathcal{L}_{(0)} &= i(\delta_C\phi_0^\dagger)\dot{\phi}_0 + i \phi_0^\dagger(\delta_C\dot{\phi}_0)
    =-i\partial_t\left(\vec{b}\cdot\vec{x}\phi_0^\dagger\dot{\phi}_0\right) = -\partial_t\left(\vec{b}\cdot\vec{x}\mathcal{L}_{(0)}\right)
\end{align}
Now consider the action given by \eqref{NLO}
\begin{align}
    \mathcal{L}_{(1)} = i\left(\phi_0^\dagger\dot{\phi_1} + \phi_1^\dagger\dot{\phi_0} - \phi_0^\dagger\s^i\partial_i\phi_0\right) 
\end{align}
Under the transformation \eqref{nlobt}, the Lagrangian transforms as
\begin{align}
    \delta_C\mathcal{L}_{(1)} &= i(\delta_C\phi_0^\dagger)\dot{\phi}_1 + i \phi_0^\dagger(\delta_C\dot{\phi}_1) + i(\delta_C\phi_1^\dagger)\dot{\phi}_0 + i \phi_1^\dagger(\delta_C\dot{\phi}_0) -i(\delta_C\phi_0^\dagger)\s^i\partial_i\phi_0 -i\phi_0^\dagger\s^i\partial_i(\delta_C\phi_0)\notag\\
    &= -i\partial_t\left(\vec{b}\cdot\vec{x}(\phi_0^\dagger\dot{\phi_1} + \phi_1^\dagger\dot{\phi_0} - \phi_0^\dagger\s^i\partial_i\phi_0)\right) - \phi_0^\dagger \vec{b}\cdot \sigma\dot{\phi}_0 + \phi_0^\dagger b^i\sigma^k\delta_{ik}\dot{\phi}_0\notag\\
    &=-i\partial_t\left(\vec{b}\cdot\vec{x}(\phi_0^\dagger\dot{\phi_1} + \phi_1^\dagger\dot{\phi_0} - \phi_0^\dagger\s^i\partial_i\phi_0)\right) = -\partial_t\left(\vec{b}\cdot\vec{x}\mathcal{L}_{(1)}\right)
\end{align}
These show that both LO and NLO actions are invariant under the transformations \eqref{lobt} and \eqref{nlobt} respectively. 

\section{Massive fermions}\label{appE}
The $c$-expansion performed in section \ref{sec3} was carried out for the massless Dirac Lagrangian. Upon including the mass term $m\bar{\Psi}\Psi$, qualitatively new features arise, depending on whether one considers the even or uneven $c$-expansion. In terms of two-component spinors, the mass term reads
\begin{align}
   \mathcal{L}^{\text{Dirac}}_{\text{mass}} = m\bar{\Psi}\Psi = im(\phi^\dagger\chi - \chi^\dagger\phi)
\end{align}
First let's look at the uneven expansion \eqref{uneven}: 
\begin{align}
    \mathcal{L}^{\text{Dirac}}_{\text{mass}} = c^{2\Delta}im\left(c^0(0) + c\left(\phi_0^\dagger\chi_0 - \chi_0^\dagger\phi_0\right) + c^2\left(\phi_0^\dagger\chi_1 + \phi_1^\dagger\chi_0 - \chi_0^\dagger\phi_1 - \chi_1^\dagger\phi_0\right) + \,...\right)\,.
\end{align}
Thus, the leading contribution appears at order $\mathcal{O}(c)$. 
In particular, there is no mass term at LO unless one rescales the mass parameter as $m \sim c^{-1}$. The first non-trivial contribution arises at NLO, and takes the standard Dirac form built from the leading spinor components $(\phi_0,\chi_0)$. This is consistent with the fact that only $\phi_0$ appears in the leading-order kinetic term.

\noindent Now, using the even c-expansion \eqref{even}, we find
\begin{align}
    \mathcal{L}^{\text{Dirac}}_{\text{mass}} = c^{2\Delta}im\left( c^0\left(\phi_0^\dagger\chi_0 - \chi_0^\dagger\phi_0\right) + c\left(\phi_0^\dagger\chi_1 + \phi_1^\dagger\chi_0 - \chi_0^\dagger\phi_1 - \chi_1^\dagger\phi_0\right) + \,...\right)\,.
\end{align}
In this case, the standard Dirac mass term already appears at leading order. 

\newpage

\bibliographystyle{jhep}
\bibliography{ref}

\end{document}